\documentclass[prx, twocolumn, superscriptaddress, longbibliography, showpacs, nofootinbib]{revtex4-1}

\usepackage{fancyhdr}
\fancyheadoffset[L,R]{\headrulewidth}
\renewcommand{\headrulewidth}{0.6pt}

\usepackage{cancel}
\usepackage[normalem]{ulem}
\usepackage{graphicx}
\usepackage{tikz,pgf}
\usepackage{color}
\usepackage{array}
\usepackage{xcolor}
\usepackage[dvips]{epsfig}
\usepackage{epsfig}
\usepackage{enumerate}
\usepackage[latin1]{inputenc}
\usepackage[T1]{fontenc}
\usepackage[hang]{subfigure}
\usepackage{bbm, dsfont}
\usepackage{amsfonts,amsmath,amssymb,stmaryrd}
\usepackage{ulem}
\usepackage{mathrsfs}
\usepackage{bbold}
\usepackage{simplewick}

\definecolor{linkcolor}{rgb}{0,0,0.6} 

\newcommand{\cA}{\mathcal{A}}
\newcommand{\cC}{\mathcal{C}}
\newcommand{\dC}{\mathscr{C}}
\newcommand{\cF}{\mathcal{F}}

\newcommand{\cO}{\mathcal{O}}
\newcommand{\cP}{\mathcal{P}}
\newcommand{\dR}{\mathscr{R}}
\newcommand{\cR}{\mathcal{R}}
\newcommand{\cS}{\mathcal{S}}

\newcommand{\qq}{\begin{equation}}
\newcommand{\qqq}{\end{equation}}
\newcommand{\al}{\begin{aligned}}
\newcommand{\all}{\end{aligned}}

\newcommand{\bfA}{{\bf A}}
\newcommand{\bfJ}{{\bf J}}
\newcommand{\bfk}{{\bf k}}

\newcommand{\bfr}{{\bf r}}
\newcommand{\bfro}{{\bf r}_1}
\newcommand{\bfrd}{{\bf r}_2}
\newcommand{\bfv}{{\bf v}}

\newcommand{\p}{\partial}

\newcommand{\A}{{\text{\tiny A}}}
\newcommand{\B}{{\text{\tiny B}}}
\newcommand{\E}{{\text{\tiny E}}}
\newcommand{\R}{{\text{\tiny R}}}

\newcommand{\bpsi}{{\mbox{\boldmath{$\psi$}}}}
\newcommand{\bchi}{{\mbox{\boldmath{$\chi$}}}}
\newcommand{\bLambda}{{\mbox{\boldmath{$\Lambda$}}}}

\newcommand{\eps}{\varepsilon}

\newcommand{\br}{\textbf{r}}
\newcommand{\bv}{\textbf{v}}

\renewcommand{\log}{\ln}
\providecommand{\avg}[1]{\left \langle #1 \right \rangle}
\providecommand{\pnt}[1]{\left ( #1 \right)}
\providecommand{\brt}[1]{\left [ #1 \right]}
\providecommand{\cur}[1]{\left\{ #1 \right\}}
\providecommand{\abs}[1]{\left| #1 \right|}
\providecommand{\df}[2]{\dfrac{#1}{#2}}
\providecommand{\f}[2]{\frac{#1}{#2}}

\begin{document}

\title{Entropy production in field theories without time reversal
 symmetry:
 \\
 Quantifying the non-equilibrium character of active matter}

\author{Cesare Nardini} 
\affiliation{DAMTP, Centre for Mathematical Sciences, University of Cambridge, Wilberforce Road, Cambridge CB3 0WA, UK}
\affiliation{SUPA, School of Physics and Astronomy, University of Edinburgh, Peter Guthrie Tait Road, Edinburgh EH9 3FD, UK}
\affiliation{Service de Physique de l'\'Etat Condens\'e, CNRS UMR 3680, CEA-Saclay, 91191 Gif-sur-Yvette, France}

\author{\'Etienne Fodor} 
\affiliation{DAMTP, Centre for Mathematical Sciences, University of Cambridge, Wilberforce Road, Cambridge CB3 0WA, UK}
\affiliation{Laboratoire Mati\`ere et Syst\`emes Complexes, UMR 7057 CNRS/P7, Universit\'e Paris Diderot, 10 rue Alice Domon et L\'eonie  Duquet,  75205 Paris cedex 13,  France}

\author{Elsen Tjhung} 
\affiliation{DAMTP, Centre for Mathematical Sciences, University of Cambridge, Wilberforce Road, Cambridge CB3 0WA, UK}

\author{Fr\'ed\'eric van Wijland} 
\affiliation{Laboratoire Mati\`ere et Syst\`emes Complexes, UMR 7057 CNRS/P7, Universit\'e Paris Diderot, 10 rue Alice Domon et L\'eonie  Duquet,  75205 Paris cedex 13,  France}%

\author{Julien Tailleur} 
\affiliation{Laboratoire Mati\`ere et Syst\`emes Complexes, UMR 7057 CNRS/P7, Universit\'e Paris Diderot, 10 rue Alice Domon et L\'eonie  Duquet,  75205 Paris cedex 13,  France}%

\author{Michael E. Cates} 
\affiliation{DAMTP, Centre for Mathematical Sciences, University of Cambridge, Wilberforce Road, Cambridge CB3 0WA, UK}


\begin{abstract}

Active matter systems operate far from equilibrium due to the continuous energy injection at the scale of constituent particles. At larger scales, described by coarse-grained models, the global entropy production rate ${\cal S}$ quantifies the probability ratio of forward and reversed dynamics and hence the importance of irreversibility at such scales: it vanishes whenever the coarse-grained dynamics of the active system reduces to that of an effective equilibrium model. We evaluate ${\cal S}$ for a class of scalar stochastic field theories describing the coarse-grained density of self-propelled particles without alignment interactions, capturing
such key phenomena as motility-induced phase separation. We show how the entropy production can be decomposed locally (in real space) or spectrally (in Fourier space), allowing detailed examination of the spatial structure and correlations that underly departures from equilibrium. For phase-separated systems, the local entropy production is concentrated mainly on interfaces with a bulk contribution that tends to zero in the weak-noise limit. In homogeneous states, we find a generalized Harada-Sasa relation that directly expresses the entropy production in terms of the wavevector-dependent deviation from the fluctuation-dissipation relation between response functions and correlators. We discuss extensions to the case where the particle density is coupled to a momentum-conserving solvent, and to situations where the particle current, rather than the density, should be chosen as the dynamical field. We expect the new conceptual tools developed here to be broadly useful in the context of active matter, allowing one to distinguish when and where activity plays an essential role in the dynamics.
\end{abstract}

\pacs{05.40.-a; 05.70.Ce; 82.70.Dd; 87.18.Gh}

\maketitle


Active matter consists of systems where energy is injected at the
level of each constituent particle, for instance to power a self-propelled
motion, before being dissipated locally~\cite{Marchetti2013RMP}. Interacting assemblies of such
particles exhibit a rich phenomenology, ranging from the transition to
collective
motion~\cite{Vicsek1995PRL,Gregoire:2004:PRL,Solon:2015:PRL,Deseigne:2010:PRL,Bricard:2013:Nature}
to the emergence of spatio-temporal chaos~\cite{Wensink:2012:PNAS} and
large-scale vortices of self-propelled composite structures~\cite{Schaller:2010:Nature,Sumino:2012:Nature}. The sustained
injection and dissipation of energy at the microscopic level drives the dynamics
out of equilibrium. Despite this, it is sometimes difficult to 
pinpoint a truly nonequilibrium signature in the emergent collective properties: the strong
microscopic departure from equilibrium does not necessarily survive in
the large scale physics. This is particularly striking in the
emergence of cohesive matter in the absence of cohesive forces through
the mechanism of motility-induced phase
separation~\cite{Tailleur:08,filyABP,Cates:15}. While clearly
non-thermal, the large-scale dynamics does not lead to steady mass
currents and closely resembles equilibrium phase separation. Accordingly, many attempts have been made to connect this
phenomenology to equilibrium
physics~\cite{Tailleur:08,Brady:2014:PRL,Yang:2014:SM,Speck:2014:PRL,Solon:2015:EPJST,Ginot:2015:PRX,Takatori:2015:PRE,Farage:2015:PRE,Marconi:2015:SM,Solon:2016:arxiv}.

The departure from equilibrium in active systems has often been studied by
introducing effective temperatures as defined by the ratio of response functions to correlators \cite{cugliandolo2011effective}. These reduce to the true temperature in the equilibrium limit because the fluctuation-dissipation theorem then holds~\cite{Loi2008PRE, Loi2011SM, Levis:15,Solon:2015:EPJST,lander2012noninvasive}. In
particular, effective temperatures have been measured experimentally,
for instance in living systems from the dynamics of injected
tracers~\cite{Mizuno, Wilhelm, gallet:2009, Visco:15,Ahmed:15b}. Generically, however, there is no direct connection between the value
of the effective temperature and the non-equilibrium nature of the dynamics encoded in the
breakdown of time-reversal
symmetry (TRS). Another characteristic feature of non-equilibrium systems is the emergence of steady-state currents, whose study has long been of interest~\cite{derrida2007,bodineau2004current,bertini2005current,bertini2015macroscopic,pilgram2003stochastic,MacKintosh:2016}. 
On the other hand, TRS breakdown in steady state is in general quantified by the global entropy
production rate ${\cal S}$ ~\cite{seifert2012stochastic}. This can be found, even far from equilibrium, directly from the probability ratio
of each realization of the dynamics to its
time reversed counterpart~\cite{lebowitz1999gallavotti, Maes1999, seifert2012stochastic,kurchan1998fluctuation}. 

The possibility that temporal or spatial coarse-graining of a nonequilibrium system can restore or partially restore TRS, creating an effective equilibrium dynamics, has been theoretically addressed in a number of different studies \cite{egolf2000,wang2016entropy,bo2014entropy,egolf2000equilibrium,cerino2015entropy,Harada:05}. 
 So far however, only a few of these address
active matter directly \cite{Tailleur:08,shim2016macroscopic,kwon2015anomalous,chaudhuri2014active,chaudhuri2016entropy,fodor2016far}. 
Our goal in the present work is to understand the
connection between emergent phenomena, such as phase
separation, and the existence of irreversibility at coarse-grained scales. We address this
question by studying the entropy production of stochastic field
theories, which describe at coarse-grained level active systems undergoing motility-induced
phase separation. 
Importantly, we progress beyond the evaluation of the global entropy production rate, which in steady state is a single number ${\cal S}$, to address the more detailed question of how this is built up of contributions from different regions of real space or reciprocal space.  This allows us to develop tools for quantifying deviations from equilibrium in a locally or spectrally resolved fashion, leading to new insight.

To capture the collective physics of self-propelled particles, various
continuum theories have been proposed based either on coarse-graining
procedures or on symmetry arguments~\cite{Marchetti2013RMP}. A
prototypical example is ``Active Model
B''~\cite{wittkowski2014scalar} which describes diffusive phase separation between
two isotropic phases of active matter at different particle densities. 
It captures in stylized form the coarse-grained 
many-body dynamics of active particles with either
discrete~\cite{Tailleur:08,Solon:2016:arxiv} or
continuous~\cite{stenhammar2013continuum,Tailleur:13} angular
relaxation of their self-propulsion direction. (Such particles can phase separate even when their interactions are purely repulsive~\cite{Cates:15}.)
Within a gradient expansion of the particle density field $\phi({\bf r})$, at zeroth order
a bulk free energy density $f(\phi)$ can be constructed, thus
providing a mapping to equilibrium~\cite{Tailleur:08}. But at square-gradient level the mapping to equilibrium is
lost and TRS is broken~\cite{stenhammar2013continuum, wittkowski2014scalar}. Active Model B thus adds minimal TRS-breaking gradient terms to Model B, which is an equilibrium square-gradient theory widely employed in the theory of critical phenomena \cite{hohenberg1977theory,chaikin2000principles}. This resembles how the KPZ equation \cite{KPZ} was constructed, by extension of the Edwards-Wilkinson model~\cite{edwards1982surface}, as a prototypical non-equilibrium model of interfacial growth.

The methods used to quantify entropy production rely on path integral representations of stochastic PDEs and 
weak-noise large deviation theory \cite{faris1982large,touchette2009large,touchettelarge,tailleur2008mapping,nardiniperturbative2016}. 
These have recently been used to address the macroscopic behaviour of diffusive systems \cite{bertini2015macroscopic,derrida2011microscopic}. In that context, several stochastic field theories associated via suitable coarse graining \cite{Dean,doi1976second,peliti1985path,janssen1976lagrangean} with lattice \cite{bertini2015macroscopic,tailleur2007mappingprl,tailleur2008mapping} and continuum models \cite{nardiniperturbative2016} have been considered, and large-deviation results for both 
 density 
\cite{nardiniperturbative2016,bertini2015macroscopic} and current \cite{derrida2007,bertini2015macroscopic,derrida2011microscopic} have been obtained. Other results on stochastic thermodynamics and fluctuation theorems for field theories have been offered~\cite{mallick2011field,leonard2013stochastic,hurtado2011symmetries,gradenigo2012entropy}.

We give in Section \ref{sec:AMB} a brief review of Active Model B. Then we define in Section \ref{sec:global}
the global entropy production rate ${\cal S}$ from the probability of observing a
realization of the fluctuating density field $\phi({\bf r},t)$. 
A success of ${\cal S}$ is to capture in a single number the importance
of non-equilibrium effects in steady state. However, a nonzero value
does not allow straightforward physical insight into where or how TRS is broken in
the coarse-grained dynamics.  To rectify this we introduce a spatial decomposition of the entropy production that allows us to locally resolve the
nonequilibrium effects of activity.  For a phase-separated state of Active
Model B, we find that the main contribution arises
from the interfaces. On this basis, one might be tempted to assume that in homogeneous phases
the nonequilibrium nature of the dynamics becomes unimportant,
yielding effective equilibrium behavior~\cite{Tailleur:08,Tailleur:13}. But in fact this holds only
at mean-field level and is broken by fluctuations, as we show, both directly, and by constructing
a spectral decomposition of the entropy
production. This decomposition generalizes to active field theories the Harada-Sasa relation (HSR)~\cite{Harada:05} 
which relates ${\cal S}$ to violations of the fluctuation-dissipation relation. The experimental feasibility to measure entropy production from HRS has already been proven~\cite{toyabe2010nonequilibrium,toyabe2007experimental, Ahmed:15c} while its application to many-body active matter systems is still elusive. We finally
extend our approach to more general systems by considering (in Section \ref{sec:AMH}) a diffusive
active density field coupled to a momentum-conserving fluid (Active Model H),
and (in Section \ref{sec:curl}) a diffusive model in which the mean current ${\bf J}$ has nonzero curl (Active Model B+). In the latter case we can choose whether to calculate ${\cal S}$ from $\phi({\bf r},t)$ as before, or from ${\bf J}({\bf r},t)$ which now contains more information. The results differ: this shows that the entropy production depends on what variables one chooses to retain in a coarse-grained description. In Section \ref{Conc} we give our conclusions.


\section{Active Model B}

\label{sec:AMB}
We consider a conserved, scalar phase-ordering system at the fluctuating
hydrodynamics level. Based on symmetry grounds, the simplest dynamics
of its order parameter field $\phi(\bfr,t)$ satisfies
\begin{equation}
  \label{eq:general_AMB}
    \dot \phi = -\nabla \cdot \bfJ , \quad   \bfJ = -\nabla \mu + {\bf \Lambda} ,
\end{equation}
where {\bf J} is a fluctuating current and ${\bf \Lambda}$
a spatio-temporal Gaussian white noise field satisfying
\begin{equation} \label{eq:AMB-noise}
  {\bf \Lambda} = \sqrt{2D}{\bf \Gamma}\;;\;\avg{ \Gamma_\alpha (\bfr, t) \Gamma_\beta (\bfr',t') } = \delta_{\alpha\beta} \delta (\bfr-\bfr') \delta (t-t') .
\end{equation}
Here
the noise strength $D$ is the ratio of the collective diffusivity to the
collective mobility; the latter has been set to unity in \eqref{eq:general_AMB}.
For systems en route to equilibrium, the deterministic part of the
current takes the form
\begin{equation}
  \bfJ_d \equiv -\nabla \mu , \quad  \mu=\mu_\E \equiv \f{ \delta \cF [\phi] }{ \delta \phi } .
\end{equation}
This is Model
B~\cite{hohenberg1977theory,chaikin2000principles}.  The chemical potential $\mu_\E$
derives from a free energy $\cF[\phi]$ which is conveniently chosen of
the $\phi^4$-type
\begin{eqnarray}
  \label{eq:AMB}  \cF[\phi] &=&\int
  \brt{ f (\phi)+ \f{\kappa}{2} | \nabla \phi |^2 } d \bfr ,
  \\
  f(\phi) &=& \f{ a_2 \phi^2 }{2} + \f{ a_4 \phi^4 }{4} .
\end{eqnarray}
Phase separation then arises, at mean-field level, whenever $a_2<0$; here $a_4$ and $\kappa$ are both positive.

For a class of phase-separating active matter models, it has been
argued
that the main effect of the activity can be captured at the fluctuating
hydrodynamic level by an additional contribution to the chemical
potential given, in its simplest form, by:
\begin{equation}\label{muAB}
  \mu=\mu_\E+\mu_\A , \quad  \mu_\A \equiv \lambda |\nabla \phi|^2 .
\end{equation}
Eqs.~(\ref{eq:general_AMB}--\ref{muAB}) define
Active Model B~\cite{stenhammar2013continuum,wittkowski2014scalar},
which is the simplest coarse-grained description of phase-separating
active systems. Note that the explicit coarse-graining of
self-propelled particles interacting via a density-dependent
propulsion speed indeed leads to a closely related dynamics, albeit with more complex
density-dependence of the various coefficients~\cite{stenhammar2013continuum,Solon:2016:arxiv}.

The defining property of $\mu_\A$ is that it cannot be written as the functional derivative of
any $\cF[\phi]$. It represents a nonequilibrium
chemical potential contribution, which violates TRS by undermining the free-energy structure of the steady
state. Interestingly, while the break-down of TRS
has often been modelled at the microscopic level with
nonequilibrium noise terms that directly break fluctuation-dissipation relations, it is the \textit{deterministic}
part the current $\bfJ$ which deviates from
equilibrium form in Active Model B. This differs from recent studies of the impact
of colored noises on Models A and B~\cite{Paoluzzi:2016:arxiv}.
Note also that the decomposition of $\mu$ into its equilibrium and nonequilibrium parts is not unique; 
since the only defining property of $\mu_\A$ is that it does not derive from a free energy, an
arbitrary equilibrium contribution can be moved into it from $\mu_\E$. This does {\em not} change any of the results of the next Section, but we will have to be more careful in Section \ref{sec:AMH} where
activity affects the stress tensor as well as the chemical potential.

\section{Entropy production}

\label{sec:global}

To quantify the breaking of TRS, the main quantity of interest is the
noise-averaged, global, steady-state entropy production rate
$\cS$. According to the precepts of stochastic
thermodynamics~\cite{sekimoto2010stochastic,seifert2012stochastic}, it is defined as~\cite{lebowitz1999gallavotti}
\begin{equation}
\label{eq:sigma} \cS = \lim_{\tau\to \infty} \cS^{\tau} ,
\quad \cS^{\tau} = \f{1}{\tau} \avg{ \log \f{ \cP[\phi] }{
    \cP^\R[\phi] } } , 
\end{equation}
where $\cP$ is the probability of a path $\{ \phi (\bfr, t) \}_{0\leq
  t\leq \tau}$, and $\cP^\R$ is the probability of its time-reversed
realization. The average $ \avg{\cdot}$ in~\eqref{eq:sigma} is taken
with respect to noise realizations. With suitable ergodicity
assumptions,\footnote{For phase-separated systems, ergodicity is
  assumed subject to specified positions for the
  interfaces between phases.\label{foot:one}} which we make throughout this paper,
averaging over one long trajectory ($\tau\to\infty$) with a single
noise realization is equivalent to the noise-average
in~\eqref{eq:sigma}. In this case the angular brackets $ \avg{\cdot} $
in \eqref{eq:sigma} can be dropped, and below we silently do this whenever it suits our purposes.
For general initial conditions $\phi(\bfr,0)$, this
single-trajectory time-average will also include a transient
contribution that scales sublinearly with its duration $\tau$, and
hence does not contribute to the entropy production {\em rate} $\cS$; the initial condition is then
irrelevant. All terms of equilibrium form ($\mu_\E$) contribute only to this transient
as we will see below. Importantly, although individual paths can have negative
entropy production, $\cS^{\tau}$ as defined in~\eqref{eq:sigma} cannot
be negative. Using standard field-theoretical
methods~\cite{Onsager:53, Martin,Dominicis,janssen1976lagrangean}, the
trajectory weight can be written as
\begin{equation}
  \label{eq:PA} \cP [\phi] = \exp \brt{ -\cA } , 
\end{equation} 
where $\cA[\phi(\bfr,t)]$ is the dynamical action.

For Active Model B as studied here, we show below that the global
entropy production can be written as
\begin{equation}
 \label{eq:Sdensity} \cS = \int \avg{ \hat \sigma (\phi, \nabla\phi, \dots ) } d \bfr ,
\end{equation}
where $\hat\sigma$ is a local function of the field $\phi$ and its
derivatives. We then
interpret the integrand $\sigma(\bfr)\equiv \langle\hat\sigma\rangle$
as a steady-state local entropy production density. Yet, such an
interpretation carries two caveats. First, there are several different
possible expressions for $\sigma({\bf r})$ which differ by transient
and/or total derivative terms, all giving the same integral
$\cS$; we return to this issue below. Second, the existence of a local entropy production density
seemingly implies additivity of $\cS$ over subsystems. This is however a subtle point which
is discussed in Appendix \ref{app:sec:local}.

The dynamical action for Active Model B
is\footnote{\label{foottwo}An equivalent expression for $ \cA [\phi] $ is
  \qq\label{eq:intro-S-nabla-1} \mathbb{A}= |\nabla^{-1}(\dot\phi +
  \nabla \cdot \bfJ_d)|^2 \qqq with the definition $\nabla^{-1}X
  \equiv \nabla \nabla^{-2}X$. The latter operation maps a scalar
  field $X$ to a vector field ${\bf Y}$ such that $\nabla.{\bf Y} = X$
  with a gauge choice $\nabla\wedge {\bf Y} = {\bf 0}$. The two forms
  for $\mathbb{A}$ in (\ref{eq:intro-S}) and
  (\ref{eq:intro-S-nabla-1}) are then related by integration by parts.
}~\cite{Onsager:53}
\begin{equation}\label{eq:intro-S} \al \cA[\phi] &= \f{1}{4D} \int
\mathbb{A} (\bfr, t) d \bfr dt ,
\\
\mathbb{A} &= - ( \dot \phi + \nabla \cdot \bfJ_d) \nabla^{-2} (\dot
\phi + \nabla \cdot \bfJ_d) .  \all
\end{equation}
Here the stochastic integral, like all those below, is defined in the
Stratonovich sense~\cite{VanKampenBook}. In Eq.~\eqref{eq:intro-S}, we
have silently omitted a time-symmetric contribution stemming from the
Stratonovich time-discretisation (see
Appendices~\ref{app:discretised} and \ref{app:discretised-strato-action} for details). The
integral operator $\nabla^{-2}$ is the functional inverse to the
Laplacian (a Coulomb integral). The action $\cA$ measures the
logarithm of the probability that a solution to
(\ref{eq:general_AMB}) is arbitrarily close to a given path
$\{\phi(\bfr,t)\}_{0\leq t\leq \tau}$.

We now introduce the time-reversal operation \begin{equation} \label{eq:MECTRS2}
\al & t \to \tau - t ,
\\
& \phi(\bfr,t) \to \phi^\R (\bfr, t) = \phi (\bfr, \tau-t) .  \all
\end{equation}
This defines $ \phi^\R (\bfr, t) $ as the path found by running a
movie of $ \phi (\bfr, t) $ backwards in time. The quantity $\cP^\R$
then represents the probability of observing the trajectory
$\phi^\R(\bfr,t)$ under the ``forward'' dynamics~\eqref{eq:general_AMB}. Since in
these dynamics the deterministic part of the current $\bfJ_d$ is a
functional of $\phi$, we have $ \bfJ_d (\bfr, t) \to \bfJ_d^\R (\bfr,
t) = \bfJ_d (\bfr, \tau-t) $, as found by
substituting~\eqref{eq:MECTRS2} into~\eqref{eq:general_AMB}. Thus the
deterministic current is not reversed, even though the actual current
$\bfJ=\bfJ_d+{\bf \Lambda }$ is equal and opposite in the forward and
backward paths, because $ \dot \phi^\R = - \dot\phi $. The forward and
backward trajectories are thus likely to require very different realizations of the
noise. Put differently, the most probable forward trajectories have
small noise contributions so that $\bfJ \simeq \bfJ_d$.  The total
current in the time-reversed trajectory, ${\bf J}^\R$, is opposed to
the deterministic one $\bfJ^\R\simeq -\bfJ_d^\R$ so that a very
atypical noise realization may be required, such that the probability of
observing the reversed trajectories is very small. 


Note that the notion of time-reversed trajectories contains a degree
of ambiguity. Here, we simply measure the probability of the noise
realization required to make the time-reversal of a trajectory $\phi(\bfr,t)$
a solution of the forward dynamics. This should
not be confused with various ``conjugacy'' operations
that instead map the forward dynamics of one system onto the backward
dynamics of another~\cite{seifert2012stochastic}. Such mappings
typically involve treating some of the parameters in the model as odd
under time reversal. This happens in magnetism, where $\cS$ can be found by comparing
a trajectory in an external magnetic field ${\bf B}$  with
the time-reversed trajectory in an external magnetic field ${-\bf
  B}$~\cite{VanKampenBook}. Such an extension of TRS is therefore mandatory to ensure the absence of entropy production in conservative magnetic systems.
In the present case, the external parameters (such as $\lambda$ in
Active Model B) must not be chosen odd under time reversal, as this
would compare forward trajectories in one system with backward
trajectories in one with different dynamics, and indeed {\em different phase
equilibria}~\cite{wittkowski2014scalar}.
Therefore, such extensions of TRS are not pertinent to the present paper.

Returning to the main issue, we now observe that the only
anti-symmetric part in the dynamical action is $\dot \phi$. The
probability for a forward trajectory to lie arbitrarily close to the
time reversed trajectory $\{\phi^\R(\bfr,t)\}_{0\leq t\leq \tau}$, as
a functional of $\{\phi(\bfr,t)\}_{0\leq t\leq \tau}$, is  $
\cP^\R = \exp \brt{ - \cA^\R } $
with \begin{equation} \label{eq:intro-SB} \al \cA^\R [\phi] &=
  \f{1}{4D} \int \mathbb{A}^\R (\bfr, t) d \bfr dt ,
  \\
  \mathbb{A}^\R &= -( \dot \phi - \nabla \cdot {\bf J}_d) \nabla^{-2}
  (\dot \phi - \nabla \cdot {\bf J}_d) , \all \end{equation} where we
have again omitted a time-symmetric contribution arising from the time
discretization. Relation (\ref{eq:sigma}) for the steady-state entropy
production can now be written
\begin{equation} \label{eq:mec1} \cS =
  \underset{\tau\to\infty}{\lim} \f{ \cA^\R - \cA }{\tau} =
  \underset{\tau\to\infty}{\lim} \f{1}{ 4 D \tau } \int (\mathbb{A}^\R
  - \mathbb{A}) d \bfr dt .
\end{equation}
Using (\ref{eq:intro-S},\ref{eq:intro-SB}), we get 
\begin{equation}
\mathbb{A}^\R-\mathbb{A} = 4 (\nabla^{-1}\dot\phi) \nabla \pnt{ \f{
    \delta \cF }{ \delta \phi } + \mu_\A } .  
\end{equation} 
Performing the spacial integral in (\ref{eq:mec1}) by parts and noting
that in the Stratonovich convention $\dot\phi\delta\cF/\delta\phi =
\dot\cF$, we find that
\begin{equation}
  \begin{aligned}  \label{eq:sigma-phi-0} 
    \cS &=
    -\lim_{\tau\to\infty} \f{1}{D \tau} \int \mu(\bfr, t) \dot \phi (\bfr,
    t) d \bfr dt\\
    &= -\lim_{\tau\to\infty} \f{1}{D \tau} \brt{ \int \mu_\A (\bfr, t)
      \dot \phi (\bfr, t) d \bfr dt + \Delta \cF } .
  \end{aligned}
\end{equation}

This result is central to all that follows. Here $ \Delta \cF = \cF
(\tau) - \cF (0) $ is the difference between the final and initial
values of the free energy functional $\cF$. This is, of course, the
only term present in the passive limit; the total free energy change
then fixes $ \tau \cS^\tau = - \Delta \cF $, so that $\cS=0$. $\Delta
\cF$ remains bounded even in active systems because $\cF(\tau)$ is
bounded below and $\cF(0)$ bounded above for sensible initial
conditions, in each case by terms linear in the system volume. Hence
this transient contribution always vanishes as
$\tau\to\infty$. Assuming ergodicity, the steady-state entropy
production then obeys \begin{equation} \label{eq:mec1a} \cS = - \f{1}{D} \int
\langle \mu_\A \dot \phi \rangle d\bfr , \end{equation} where the average
$\avg{\cdot}$ is now taken over the stationary measure. Importantly (see footnote~\ref{foot:one})
since we are interested in phase separation and similar situations of
broken translational symmetry, we should understand the angle brackets
in this expression as being noise-averages within an ensemble where
interfaces between phases
are held stationary in time. We can then define a local steady-state entropy
production density 
\begin{equation} \label{eq:sigma-phi} \sigma_\phi (\bfr) \equiv -
  \f{1}{D} \langle \mu_\A \dot \phi \rangle (\bfr) ,
\end{equation} 
whose integral is the entropy production:
\begin{equation} \label{eq:mec2} \cS = \cS_\phi \equiv \int \sigma_\phi (\bfr) d \bfr .  \end{equation}
Here and in \eqref{eq:sigma-phi} we have added a subscript $\phi$ to distinguish
these forms from expressions appearing later on (Section \ref{sec:curl}) in which the current ${\bf J}$ rather
than the density field $\phi$ is used to define trajectories.

As already mentioned following \eqref{eq:Sdensity} above, there are several other local quantities which have the same integral, $\cS$, and hence have equal claim to be called the local entropy production density. 
Indeed, 
\begin{equation} \label{eq:variants} -\sigma_\phi D \equiv
  \langle{\mu\dot\phi}\rangle \,,\,\langle{\mu_\A\dot\phi}\rangle \,,\, \langle{{\bf J}.\nabla\mu_\A}\rangle
 \,,\, -\langle{{\bf J}.{\bf J}_d}\rangle
\end{equation} 
are all equivalent for the purposes of computing ${\cal S}$. The first two forms differ by a transient contribution $\Delta {\cal F}/\tau \to 0$, as do the last two. The latter pair are found from the former by partial integration, differing from them by terms of the form $\nabla.{\bf \Upsilon}_{d,\A}$ where ${\bf \Upsilon}_d = {\bf J}\mu$ and ${\bf \Upsilon}_\A = {\bf J}\mu_\A$. Our numerical studies  suggest that these alternative candidates for local entropy production are practically indistinguishable in the case of Active Model B. A more complex situation arises  for Active Model B+, as described in Section \ref{sec:curl}.


\begin{figure*}
  \centering
  \includegraphics[width=.3\linewidth]{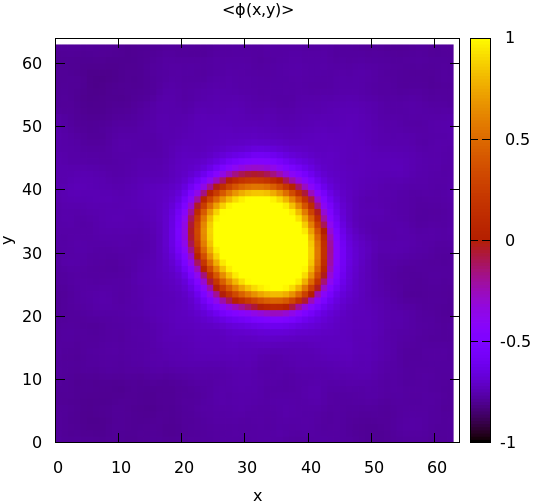}  \includegraphics[width=.3\linewidth]{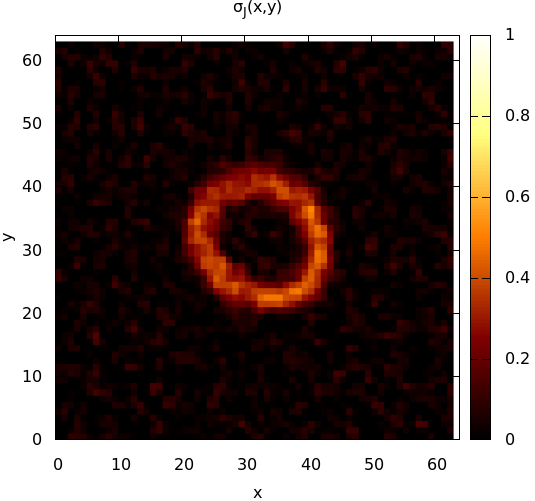}  \raisebox{.1cm}{\includegraphics[width=.38\linewidth]{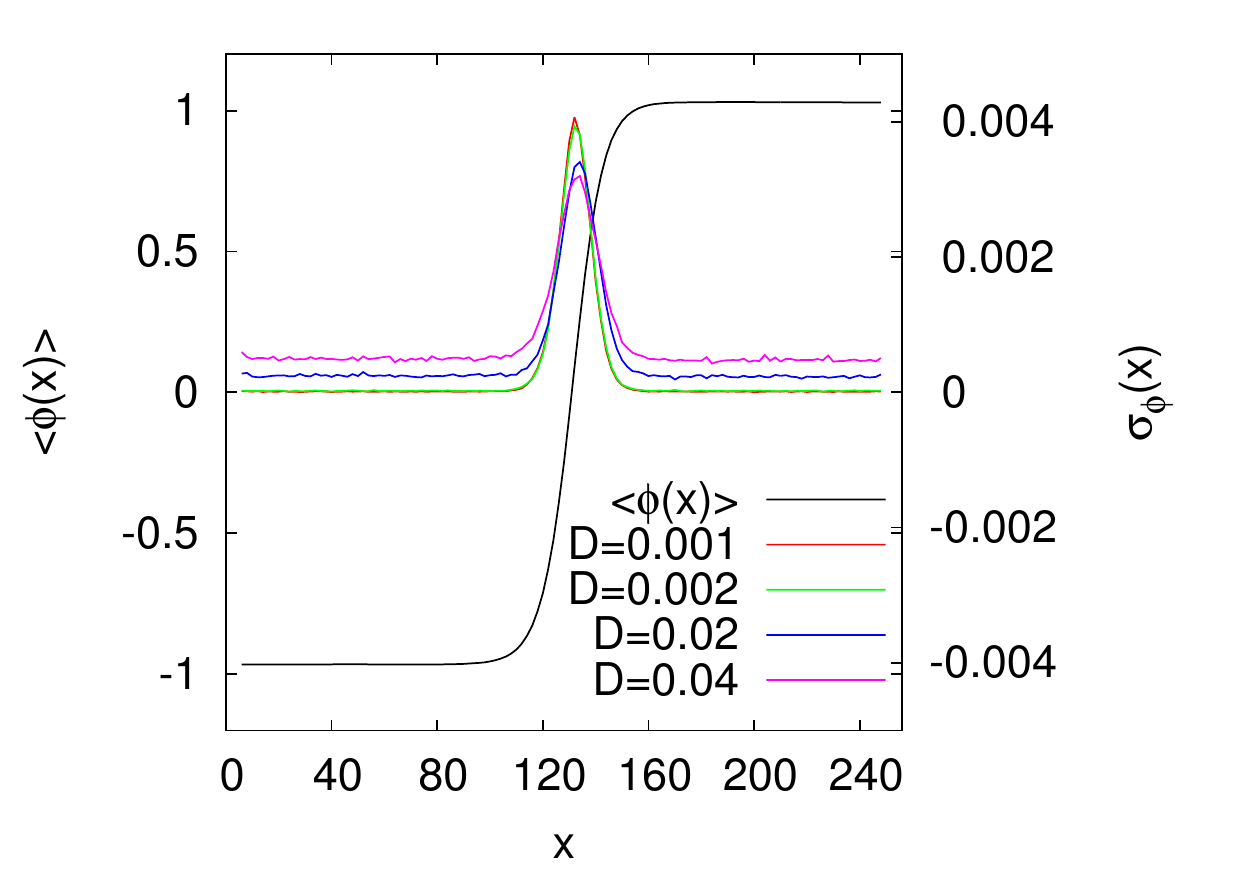}}
  \caption{\label{fig:sigmarealspace}
    {\bf Left}: Density map of a fluctuating phase-separated droplet in Active Model B in 2D. {\bf Center:} Local contribution to the entropy production $\sigma({\bf r})$ showing a strong contribution at the interfaces. {\bf Right:} Density and entropy production for a 1D system comprising a single domain wall, for various noise levels $D\ll a_2^2/4a_4$. The entropy production is strongly inhomogeneous, attaining a finite value as $D\to0$ at the interface between dense and dilute regions and converging to zero in the bulk in this limit. Values of the parameters used are $a_2=-0.125$, $a_4=0.125$, $\kappa=8$, and $\lambda=2$.
		}
\end{figure*}

\subsection{Spatial decomposition of entropy production}

\label{sec:AMB-real_space}

We now turn to the numerical evaluation of $\sigma_\phi(\bfr)$ in
Active Model B. We defer to future work any study of the critical region of the model, focusing
instead on the weak noise limit (small $D$) where sharp interfaces
form between high and low density phases, respectively denoted by
$\phi_h>0$ and $\phi_l<0$. Note that $\phi_h \neq -\phi_l$ unless
$\lambda = 0$, since activity breaks $\pm\phi$ symmetry in
$\cF$~\cite{wittkowski2014scalar}.  To study phase separated states in one dimension
we
consider a single domain wall in the centre of the system
and impose $\nabla\phi = 0$ on the distant
boundaries. We use finite difference methods with mid-point spatial
discretisation to integrate (\ref{eq:general_AMB}) via a fully explicit first order
Euler algorithm. Importantly, the discretised system exactly respects
detailed balance whenever $\lambda=0$, as shown in
Appendix~\ref{app:discretised} where further numerical details are
given. For numerical purposes we evaluate the entropy production
density as 
\begin{equation} \label{eq:sigma-phi-ito} \sigma_\phi (\bfr) =
-\lim_{t\to\infty} \f{1}{D t} \int \mu_\A (\bfr, t) \nabla^2 \mu
(\bfr, t) d t ,
\end{equation}
whose equivalence with~\eqref{eq:sigma-phi} is established, within our
discretization scheme, in Appendix~\ref{app:discretised-ito-EP}.

Our numerics in both one and two dimensions show that the local entropy production density
$\sigma_\phi({\bf r})$ is strongly inhomogeneous, being large at the
interface between phases, but small within these phases
(Fig.~\ref{fig:sigmarealspace}). To quantitatively explain this, we consider the weak noise expansion
of the density
\begin{equation}
  \label{eq:AMB-phi-decomposition} \phi = \phi_0 +
  \sqrt{D} \phi_1+ D\phi_2 + \cO (D^{3/2}) .
\end{equation}
In the weak noise limit, standard field-theoretical methods, which we outline in Appendix~\ref{app:small_noise}, show the
dynamics of $\phi_0$ and $\phi_1$ reduce to: 
\begin{eqnarray}
  \dot \phi_0 &=& - \nabla \cdot {\bfJ}_d(\phi_0)\label{eq:dynphi0}\\
  \dot \phi_1 &=& \nabla^2 \brt{
  \f{\delta \cF_0}{\delta \phi_1} + 2\lambda  \nabla \phi_0 \cdot \nabla \phi_1 } + \nabla \cdot {\bf \Gamma}\label{eq:dynphi1}
\end{eqnarray}
where ${\bf \Gamma}$ is a standard Gaussian white noise as in~\eqref{eq:AMB-noise}
and
\begin{equation} \label{eq:F_0} \cF_0 = \int \brt{ ( a_2 + 3 a_4 \phi_0^2 )
  \f{\phi_1^2}{2} + \kappa \f{|\nabla \phi_1|^2}{2} } d\bfr .  
\end{equation}
This shows, as expected, that the statistics of $\phi_{0,1}$ are
independent of $D$ at leading order.

\begin{figure}
	\centering
	\includegraphics[width=\linewidth]{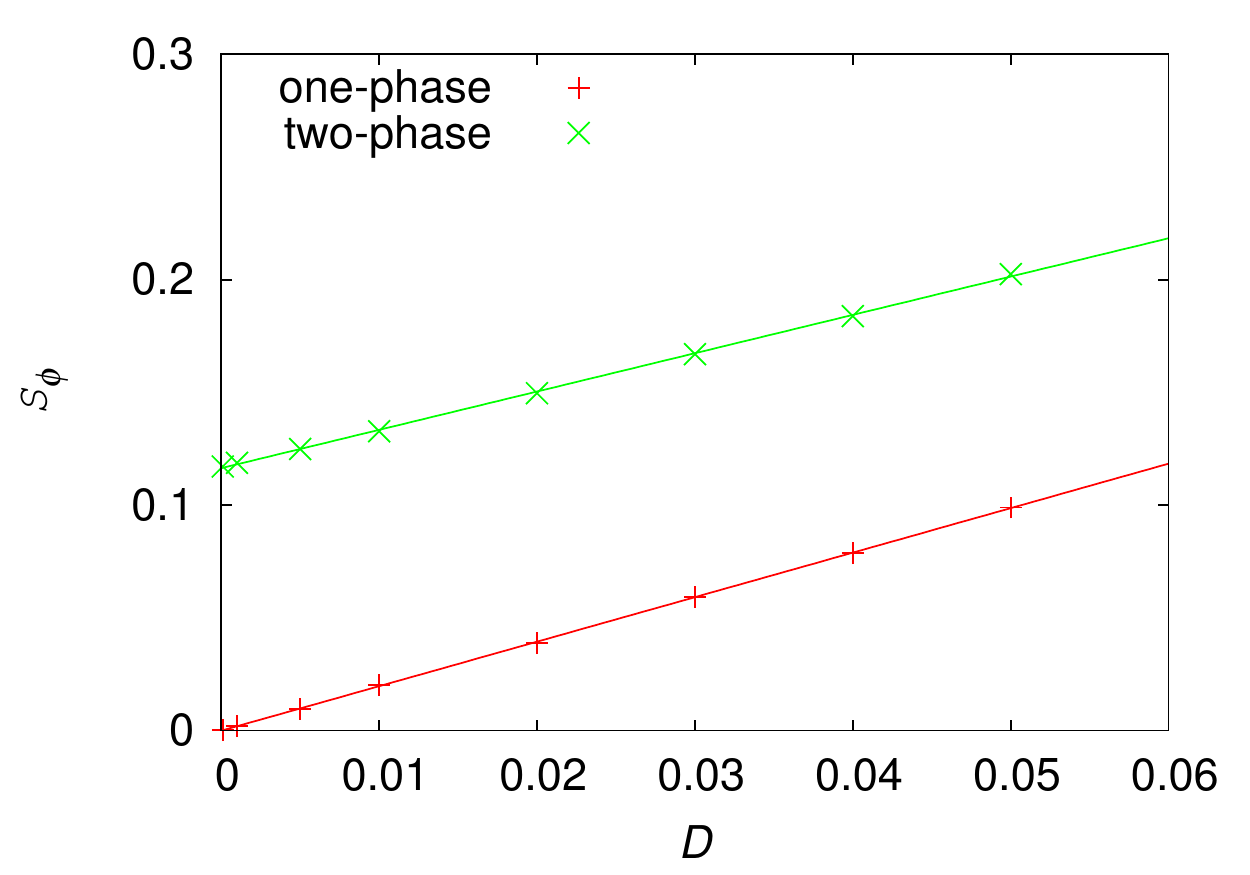}
	\caption{\label{fig:scalingD} Total entropy production
          $\cS_\phi$ for Active Model B in one dimension for the case
          where the system consists of a single uniform phase
          ($a_2>0$, $a_4>0$, red) and for the case where it shows
          coexistence between a high-density and a low density phase
          ($a_2<0$, $a_4>0$, green). (Parameter values
          are $a_2=\pm0.25$ for the uniform $(+)$ and phase-separates $(-)$ states, with $a_4=0.25,\kappa=4,\lambda=1$.) In the two-phase system, any putative sub-extensive (interfacial) term scaling as $\sqrt{D}$ would be swamped by the extensive $D^1$ contribution from bulk phases.
           }
\end{figure}

We first consider the case where the mean-field dynamics for $\phi_0$
has relaxed to a constant profile. In this case, it follows from
(\ref{eq:sigma-phi}) that
\begin{equation} \label{eq:EP-AMB-homo} \sigma_{\phi}(\bfr) = -
  \lambda \sqrt{D} \langle |\nabla \phi_1|^2 \dot \phi_1 \rangle + \cO
  (D).  
\end{equation}
Inspection of Eq.~\eqref{eq:dynphi1} shows that the only TRS-breaking
term is proportional to $\nabla \phi_0$ so that, in a homogeneous
state, $\phi_1$ has an equilibrium dynamics controlled by the
free-energy ${\cal F}_0$. The latter is even in $\phi_1$ so that the
term of order $\sqrt{D}$ in the expansion~\eqref{eq:EP-AMB-homo} of
$\sigma_\phi$ vanishes by symmetry. We conclude that in bulk phases
$\sigma_{\phi} \propto D$. We have checked this by simulations in
single-phase systems where the total entropy production is indeed
shown to scale as $\cS\propto D$; see Fig.~\ref{fig:scalingD}.


In contrast however, Fig.~\ref{fig:scalingD} also shows that the total
entropy production $\cS$ for a phase-separated system remains
finite as $D\to 0$. This $D^0$ term dominates at small enough $D$ and
is caused by entropy production localised near the interfaces between
phases. This is graphically confirmed in the spatial map of the local
entropy production density $\sigma_{\phi}({\bf r})$ shown in
Fig.~\ref{fig:sigmarealspace} which shows that, as $D\to 0$,
$\sigma$ vanishes aways from interfaces but has a finite contribution
in their vicinity.
This can be understood from the
expansion~\eqref{eq:AMB-phi-decomposition} where $\phi_0(\bfr)$ is now
a (steady) phase-separated solution of the mean-field
dynamics~\eqref{eq:dynphi0}, with a prescribed value of $\int
\phi(\bfr) d\bfr$. Inserting~\eqref{eq:AMB-phi-decomposition} in the
definition~\eqref{eq:sigma-phi} of $\sigma_\phi$ gives
\begin{equation}
  \al
  \sigma_\phi(\bfr) &=  - \f{\lambda}{\sqrt{D}} |\nabla\phi_0|^2 \langle \dot\phi_1\rangle - \lambda |\nabla\phi_0|^2 \langle \dot\phi_2\rangle 
  \\
  & \quad - 2\lambda \nabla \phi_0 \cdot \langle \nabla \phi_1 \dot \phi_1 \rangle + \cO (\sqrt{D}) .
  \all
\end{equation}
In the steady state, $\langle \dot \phi_i\rangle = \partial_t \langle
\phi_i \rangle=0$ so that
\begin{equation}\label{eq:EP-AMB-PS}
\sigma_{\phi}(\bfr) = 
-2\lambda \nabla \phi_0 \cdot
\langle
\dot{\phi}_1 \nabla \phi_1
\rangle + \mathcal{O}(\sqrt{D})
\end{equation}
where the average is taken over the stationary measure of the
dynamics~\eqref{eq:dynphi1} for $\phi_1$. 
Given that the statistics of $\phi_1$ are, by
construction, $D$-independent, it follows that $\sigma_\phi(\bfr) =
\cO (D^0)$ wherever the gradient of the deterministic solution is
finite, \textit{i.e.}, at interfaces.

\begin{figure}
  \centering
  \includegraphics[width=\linewidth]{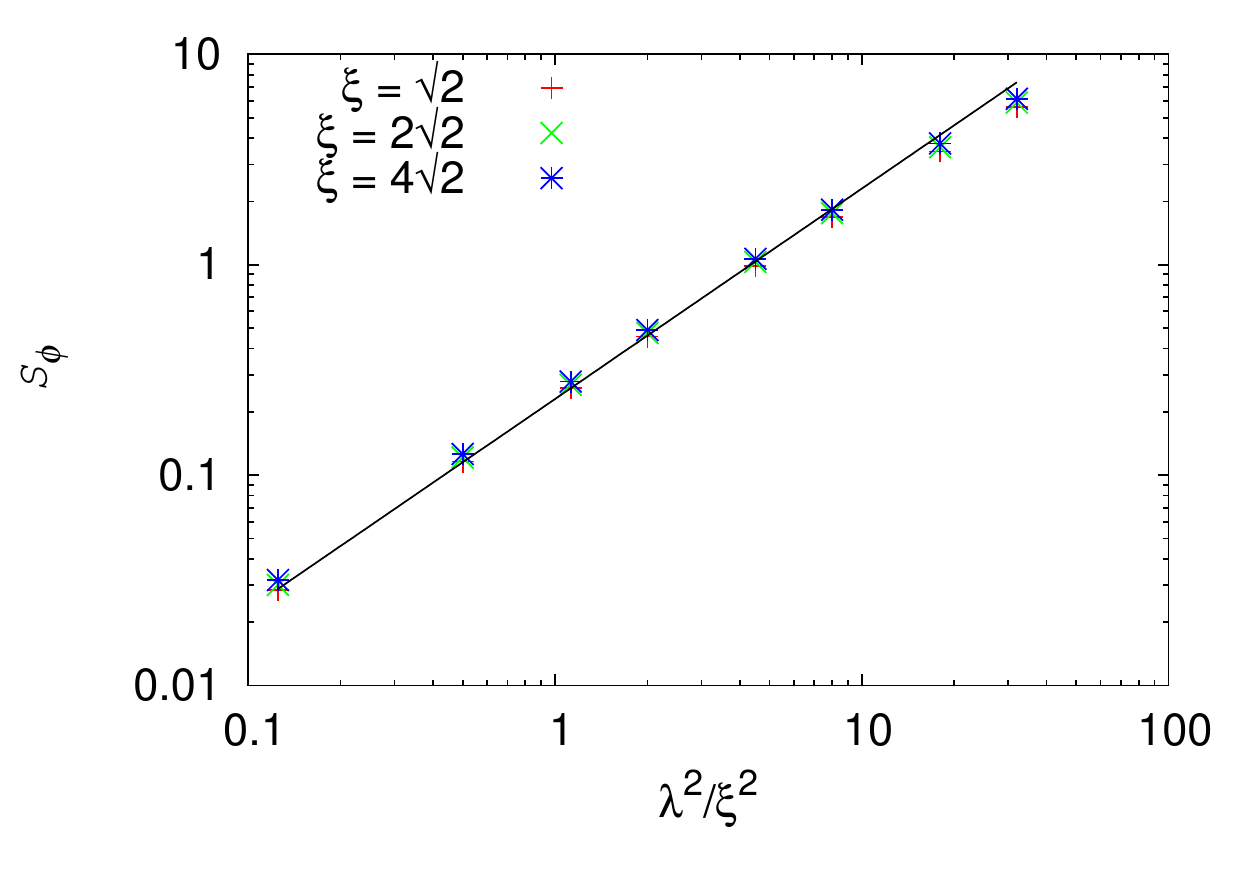}
  \caption{\label{fig:scalinglambda}
	Scaling of the total entropy production as a function of $\lambda$, showing that $\cS_\phi\propto\lambda^2$, as predicted analytically. Here $\xi=\sqrt{-\f{\kappa}{2a_2}}$ is the passive interface width. Values of the parameters used are $(a_2=-0.5,a_4=0.5,\kappa=2,D=0.001)$, $(-0.25,0.25,4,0.001)$ and $(-0.125,0.125,8,0.001)$ for red, green and blue points respectively.
}
\end{figure}

In Fig.~\ref{fig:scalinglambda}, we show the dependence of the total
entropy production $\cS_\phi$ on $\lambda$; this scales as $\cS_\phi\propto
\lambda^2$. No linear term is possible, because it would
mean that of two different Active Model B systems, with other
parameters the same but with opposite signs of $\lambda$, one would
have negative global entropy production in steady state.
Accordingly in~\eqref{eq:EP-AMB-PS}, the term $\langle
\dot{\phi}_1 \nabla \phi_1
\rangle$ must itself be of order $\lambda$ in general. This explains the
quadratic scaling, but only to leading order in small $\lambda$ (and $D$).
In practice we find this scaling over a wide range of $\lambda$ at small $D$ (Fig.~\ref{fig:scalinglambda}),
and also at larger $D$ (not shown) but we have no explanation for these results at present.

In summary, we have found above that $\sigma_\phi
\sim D^0$ where $\langle \nabla \phi \rangle \neq 0$ whereas $\sigma_\phi\sim D$ in
bulk phases (Fig.~\ref{fig:scalingD}). This confirms quantitatively that the
TRS breakdown induced by activity plays its largest role at
interfaces~\cite{wittkowski2014scalar,Solon:2016:arxiv}. More generally, the results exemplify how the local entropy production density $\sigma({\bf r})$ can provide a good quantitative tool to understand the breaking of TRS when the system is inhomogeneous. This is likely to be important in situations more complicated than  phase separation, including those where activity parameters such as swim-speed vary in space (see~\cite{stenhammar2016} and also Section~\ref{sec:curl}). In such cases analytical progress could be more difficult, but the numerical evaluation of $\sigma_\phi({\bf r})$ should remain tractable. 
On the other hand, this approach gives little information about the character of TRS breaking in homogeneous phases whose translational invariance forces $\sigma_\phi(\bfr)$ to be simply flat. In such cases spatial structure is captured by correlation functions rather than mean values; we address this next. 

\subsection{Spectral decomposition of entropy production from
  fluctuation and response}

\label{sec:AMB-HS}

The fluctuation-dissipation theorem (FDT) is a fundamental property of
equilibrium dynamics~\cite{Kubo}. It states that the response to an
external perturbation is entirely determined by spontaneous
fluctuations \textit{in absence} of the perturbation. For active field
theories, and non-equilibrium systems more widely, there is no
general relation between correlation and response. Thus, violation
of the FDT in a coarse grained description at some scale is a proof
that activity matters dynamically at this scale, since no passive
model could ever give rise to such a violation. In equilibrium, the
ratio of correlators to response functions is set by the bath temperature. For
nonequilibrium dynamics, an effective temperature is sometimes
introduced by analogy with the FDT by looking for asymptotic regimes (in time or frequency) where constant values arise for the correlator/response ratio~\cite{Kurchan:13,Levis:15}. One drawback of such
a definition of the effective temperature is that it generically is different for every
perturbation and observable under scrutiny, and therefore bears no universal meaning.

In this Section, we provide a quantitative relation between the rate
of entropy production $\cS$ and the FDT violation for the stochastic field
dynamics governed by~\eqref{eq:general_AMB}. We then present detailed numerical
results for the case of Active Model B. For an overdamped
single particle driven out of equilibrium by an external non-gradient
force, such a connection was discovered by Harada and
Sasa~\cite{Harada:05}. They showed that the violation of the FDT
relating the correlation of the particle position and the response to
a constant force provides a direct access to the entropy production
rate. Therefore, one important success of the Harada-Sasa relation (HSR) was
to identify, among all possible violations of the FDT in a
non-equilibrium system, the particular observable and perturbation that
quantitatively determine $\cS$. The HSR was later generalised to other types of equilibrium
dynamics, including the under-damped case and that with temporally correlated
noise~\cite{deutsch2006energy}. HSRs have also been
developed for systems with a time-scale separation between fast and slow
degrees of freedom~\cite{wang2016entropy}, and for the density field
in a system of particles driven by a time-independent external force
field~\cite{harada2009macroscopic}. In active matter, some of us
recently announced a generalisation of the HSR in a
non-equilibrium microscopic model of particles propelled by correlated
noises~\cite{fodor2016far}.

We consider the correlation function $C$ and the response
$R$ to a change of the chemical potential $\mu \to \mu - h$:
\begin{equation} \label{eq:AMB-HS-C-R}
	\al
		C(\bfro,\bfrd,t-s) &\equiv \avg{ \phi(\bfro,t) \phi(\bfrd,s) } ,
		\\
		R(\bfro,\bfrd,t-s) &\equiv \left. \frac{\delta \langle \phi(\bfro,t)\rangle }{\delta h(\bfrd,s)}\right|_{h=0} .
	\all
\end{equation}
Before outlining the derivation of our generalized HSR, we give the result:
\begin{equation} \label{eq:HS-AMB}
	\al
		\cS &= \int \sigma_\phi(\bfk,\omega) d\bfk d\omega ,
		\\
		\sigma_\phi (\bfk, \omega) &\equiv \f{ \omega k^{-2} }{(2\pi)^{d+1} D} \brt{ \omega C(\bfk,  \omega) - 2 D R(\bfk,\omega) } ,
	\all
\end{equation}
where
\begin{equation}
	\al
		C(\bfk, \omega) & \equiv \int C(\bfro,\bfrd,t) e^{i\bfk\cdot (\bfro-\bfrd) + i\omega t} d\bfr_{1,2} dt ,
		\\
		R(\bfk, \omega) &\equiv \int R(\bfro,\bfrd,t) e^{i\bfk\cdot (\bfro-\bfrd)} \sin(\omega t) d\bfr_{1,2} dt .
	\all 
\end{equation}
The spectral decomposition of the entropy production
rate~\eqref{eq:HS-AMB} expresses $\cS$ as an integral over Fourier
modes of a spectral density $\sigma_\phi(\bfk,\omega)$. This density
can formally be seen as the contribution to $\cS$ of the modes in
$[\bfk,\bfk+d\bfk]\times[\omega,\omega+d\omega]$. It vanishes for thermally
equilibrated modes, as enforced by the FDT: $ \omega C (\bfk, \omega)
= 2 D R (\bfk, \omega) $. The spectral decomposition is particularly
interesting for uniform systems, where the real-space local density
$\sigma_\phi(\bf r)$ defined in Eq.~\eqref{eq:sigma-phi} is obliged to
be constant by translational invariance. It is natural in such systems
to instead consider entropy production as a function of wavevector (and
perhaps also frequency), and~\eqref{eq:HS-AMB} provides the appropriate tool for
doing so. Performing the $\omega$ integral allows one to quantitatively explore whether activity
matters dynamically at a given spatial scale. A further asset of the
generalized HSR is that it allows one to evaluate $\cS$ without complete information about the equations of motion. In particular, as detailed below, we assumed these to lie in the broad class defined by~\eqref{eq:general_AMB} but made no assumption about the form of the nonequilibrium chemical potential $\mu_\A$.
This is in
contrast to the real-space methods presented in
Sec.~\ref{sec:AMB-real_space}.

We now give the derivation of the HSR result. In terms of the dynamic
action $\cA^h$ of the perturbed dynamics, using the property $ \delta
\cP = - \cP \delta \cA^h $, the response can be written as
\begin{equation} \label{eq:AMB-response-implicit}
	R (\br_1,\br_2, t) = - \avg{ \phi (\br_1, t) \left. \df{ \delta \cA^h }{ \delta h (\br_2, 0) } \right|_{h=0} }  .
\end{equation}
We need $\cA^h$ only to first order in $h$:
\begin{equation}
	\delta \cA^h = - \f{1}{2D} \int h \pnt{ \p_t \phi - \nabla^{2} \mu } d \bfr dt + \cO \pnt{ h^2 } ,
\end{equation}
so that the spatially diagonal response obeys
\begin{equation}
	R (\br,\br, t) = \f{1}{2D} \avg{ \phi (\br, t) \pnt{ \p_t \phi - \nabla^{2} \mu } (\br, 0) }  .
\end{equation}
The time-antisymmetric part follows as
\begin{equation}
	\al
		R (\br, \br, t) &- R (\br, \br, -t) = - \f{1}{D} \p_t C (\br,\br, t) -
		\\
		&- \f{1}{2D} \nabla^{2}_{(2)} \avg{ \phi (\br, t) \mu (\br, 0) - \phi (\br, 0) \mu (\br, t) } ,
	\all
\end{equation}
where we have used $ C(\bfr,\bfr,t) = C(\bfr,\bfr,-t)$
via~\eqref{eq:AMB-HS-C-R}. Here $\nabla^{2}_{(2)}$ denotes the
Laplacian operator acting on the second spatial variable of the
function on which it acts. Comparing with~\eqref{eq:mec1a} and
ignoring subdominant boundary terms, we deduce
\begin{equation} \label{eq:app:HS-AMB-real}
	\al
		\cS = \f{1}{D} \underset{t\to0}{\lim} \int \p_t \nabla^{-2}_{(2)} & \cur{ D \brt{ R (\br, \br, t) - R (\br,\br, -t) } \right.
		\\
		& \left. + \p_t C (\br, \br, t) } d\bfr d t .
	\all
\end{equation}
This is our generalised Harada-Sasa relation in real space and
time. Fourier transforming then yields~\eqref{eq:HS-AMB}.

We have computed numerically the spectral decomposition of the entropy
production rate for Active Model B, using periodic boundary
conditions. We extracted the response using the algorithm
of~\cite{Villamaina:08} where two systems, one with unperturbed and
the other with perturbed chemical potential, are driven by the same
noise realisation. This allows a very precise measurement of the
response, even without averaging over noise realizations. Nonetheless, a
separate simulation for each value of $\bfk$ and $\omega$ is needed in order to
compute the response $ R (\bfk,\omega)$. In contrast, a single run
(long enough to ensure good statistical averaging) is sufficient to
measure the correlator $C(\bfk, \omega)$ for all $\bfk$ and $\omega$ simultaneously.
This is why our statistical accuracy is better for $C(\bfk,\omega)$ than for $R(\bfk,\omega)$.

\begin{figure}
\includegraphics[width=\linewidth]{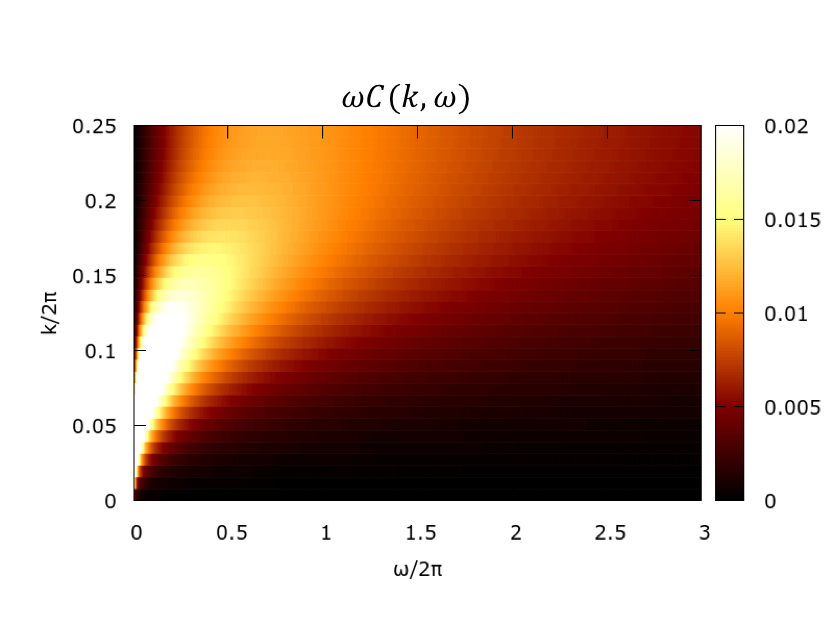}
\includegraphics[width=\linewidth]{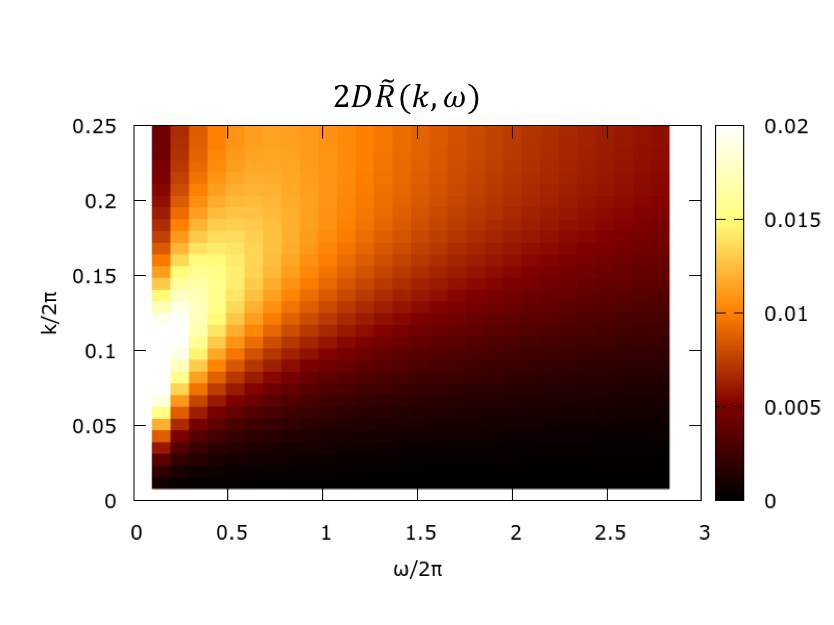}
\caption{\label{fig:figHS-1}
	(a)~Correlation $\omega C(\bfk, \omega)$ and (b)~response $2 D R(\bfk, \omega)$ for Active Model B in one dimension, found by numerical simulations as detailed in the main text. Parameter values: $-a_2=a_4 = 0.25$,  $k=4$, $D=0.05$ and $\lambda=8$.
} 
\end{figure}

\begin{figure}
\includegraphics[width=\linewidth]{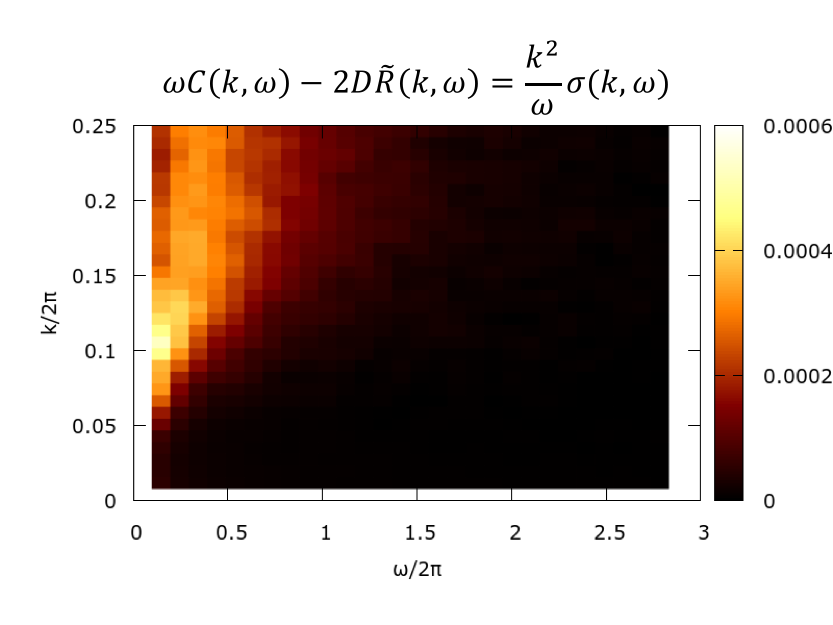}
\includegraphics[width=\linewidth]{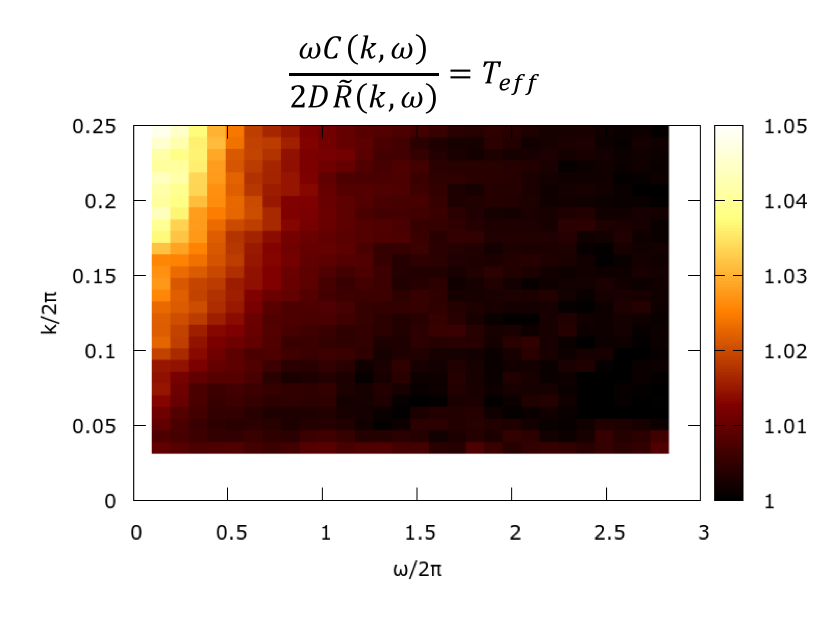}
\caption{\label{fig:figHS-2}
	(a)~Correlator-response mismatch $\omega C -2D \tilde R$ $\propto \sigma_\phi({\bf k},\omega)k^2/\omega$ and (b)~normalised effective temperature $ {\cal D}({\bf k},\omega)/D$ as a function of ${\bf k}$ and $\omega$ for Active Model B in one dimension. Parameter values as in Fig.~\ref{fig:figHS-1}. (Control simulations for the passive limit $\lambda = 0$ recover a statistically insignificant signal for both plots, not shown, confirming that our numerical algorithm respects TRS when present.)
}  
\end{figure}

In Fig.~\ref{fig:figHS-1}, we plot both quantities in a
single-phase system for a wide range of wavelengths and
frequencies. Their mismatch spectrally quantifies the breakdown of TRS
via the entropy production, as shown in Fig.~\ref{fig:figHS-2}(a). In
Fig.~\ref{fig:figHS-2}(b), we present an `effective temperature',
defined through $ {\cal D}(\bfk,\omega)/D = \omega C (\bfk,\omega) / 2
D R (\bfk,\omega)$. This is a popular measure of FDT violations (see above)
whose shortcomings are apparent here. In particular, while the
effective temperature is peaked at large wavenumbers, this is true
neither of the FDT mismatch nor of $\sigma_\phi({\bf k},\omega)$
itself. The additional factor of $\omega/k^2$ relating the FDT
mismatch to $\sigma_\phi({\bf k},\omega)$ makes the latter highly
sensitive to both high-frequency and low-wavenumber information. We
have checked that the total entropy production $\cS = \int
\sigma_\phi({\bf k},\omega) d\bfk d\omega$ nonetheless approaches numerically the
same quantity calculated in real space via
\eqref{eq:sigma-phi-ito}. The spectral density plot in
Fig.~\ref{fig:figHS-2}(a) does therefore include most of the relevant
frequency and wavenumber range. Deferring fuller investigations to
future work, we can conclude that our generalized HSR
for Active Model B yields a nontrivial spectral decomposition of the
entropy production, including features not captured by the effective
temperature paradigm.

Note that measuring the probability of trajectories and their time-reversed counterparts is difficult to do in simulations and even more so in experiments. On the other hands, physicists have an established expertise in measuring FDT both in simulations and experiments. Eq. (\ref{eq:HS-AMB}) provides a direct link for such measurements to entropy production that waits exploration for active matter systems.

\section{Active Model H}

\label{sec:AMH}

We now turn attention to Active Model H in which the diffusive dynamics of $ \phi (\bfr, t)$ is coupled to a momentum-conserving fluid with velocity $ \bfv (\bfr, t)$. This model was first introduced in~\cite{tiribocchi2015active} to address the role of fluid motion in active phase separation. In that work, it was shown that for so-called contractile activity (where particles pull fluid inward along a polar axis, determined in effect by $\nabla\phi$, and expel it equatorially) domain growth can cease at a length scale where diffusive coarsening is balanced by the active stretching of interfaces. As a generalisation to active systems of the well known Model H~\cite{hohenberg1977theory}, this model offers an interesting arena in which to explore coupling of an active scalar to a second field whose dynamics is dissipative, but which would obey TRS without coupling to the active field.

Diffusive dynamics now takes place in the frame of the moving fluid so that \eqref{eq:general_AMB} acquires an advective term,
\begin{equation} \label{eq:AMH-phi}
	\dot \phi + \bfv \cdot \nabla \phi = -\nabla \cdot \bfJ ,
\end{equation}
where we retain from Active Model B the equations for the diffusive current
\begin{equation} \label{eq:MECAMH2}
	\bfJ = \bfJ_d + \bLambda , \quad \bfJ_d = -\nabla ( \mu_\E + \mu_\A ) .
\end{equation}
We assume the fluid is incompressible and of unit mass density. The thermal Navier-Stokes equation for momentum conservation then reads
\begin{equation} \label{eq:AMH-NS}
	(\p_t  + \bfv \cdot \nabla ) \bfv = \eta \nabla^2 \bfv - \nabla p + \nabla \cdot  (\mathbf{\Sigma}+\mathbf{\tilde\Gamma}) ,
\end{equation}
with the noise stress specified by~\cite{chaikin2000principles}
\begin{equation} \label{eq:Gamma}
	\al
		&\avg{\tilde \Gamma_{\alpha\beta} (\br ,t) \tilde\Gamma_{\gamma\nu} (\br', t') } =
		\\
	 	& \quad 2 \eta D \pnt{ \delta_{\alpha\gamma}  \delta_{\beta\nu} + \delta_{\alpha\nu}  \delta_{\beta\gamma} } \delta \pnt{ \br - \br' } \delta \pnt{ t - t' }  .
	\all
\end{equation}
(Recall that we set the mobility of the $\phi$ field to unity so that $D = k_BT$.)
In~\eqref{eq:AMH-NS}, the pressure field $p({\bf r},t)$ subsumes all isotropic stress contributions and enforces fluid incompressibility ($\nabla \cdot \bfv = 0$). The deviatoric stress tensor $\mathbf{\Sigma}$ is traceless and symmetric.

In passive systems, $\mathbf{\Sigma}=\mathbf{\Sigma}^\E$ can be derived from a free-energy functional $ \cF [\phi] $ by standard procedures~\cite{chaikin2000principles}; restoring isotropic pressure contributions one finds
\begin{equation}	\label{eq:AMH-Sigmaeq}
	\nabla \cdot \mathbf{\Sigma}^\E = - \phi \nabla \mu_\E .
\end{equation}
For an active system, the relation between the deviatoric stress tensor and the free energy breaks down, taking instead the more general form
\begin{equation} \label{eq:AMH-Sigma}
 \mathbf{\Sigma} = \mathbf{\Sigma}^\E + \mathbf{\Sigma}^\A ,
\end{equation}
where $\mathbf{\Sigma}^\E$ obeys~\eqref{eq:AMH-Sigmaeq}. Note that even if activity creates a chemical potential contribution that is of equilibrium form (as arises in MIPS to zeroth order in gradients), this {\em does not}
lead to a relation $\nabla\cdot\mathbf{\Sigma}^\A = -\phi\nabla\mu_\A$ as one might expect from~\eqref{eq:AMH-Sigmaeq} ~\cite{tiribocchi2015active}. This means it is no longer possible, as it was in Active Model B, to shift some arbitrary part of $\mu_\E$ into $\mu_\A$ without changing the results. Conversely, one cannot declare all terms in $\mu$ that are of equilibrium form to be part of $\mu_\E$ as we did for Active Model B in Section \ref{sec:AMB}; we must now be careful to include in $\mu_\A$ any chemical potential contribution that does not contribute to the stress via~\eqref{eq:AMH-Sigmaeq}. For this reason, when an external perturbation $\mu\to\mu-h({\bf r})$ is used to calculate response functions below, we will need to specify whether it is $\mu_\A$ or $\mu_\E$ that is being perturbed. Note also that applying a perturbation $h({\bf r})$ to the coarse grained theory need not be equivalent to applying a similar one at microscopic level; the latter would cause more complicated shifts in both $\mu^\E$ and $\mu^\A$~\cite{Solon:2016:arxiv}.

Active Model H comprises the equations (\ref{eq:AMH-phi}-\ref{eq:AMH-Sigma}). In~\cite{tiribocchi2015active}, it was further assumed that $\cF[\phi]$ is given by~\eqref{eq:AMB}, from which the deviatoric equilibrium stress $\mathbf{\Sigma}^\E$ takes a dyadic gradient form. This form turns out to be shared by $\mathbf{\Sigma}^\A$ to leading order in gradients:
\begin{equation} \label{eq:MECstress1}
	\mathbf{\Sigma}^\E = - \kappa (\nabla\phi)(\nabla\phi) , \quad \mathbf{\Sigma}^\A = - \zeta (\nabla\phi)(\nabla\phi) .
\end{equation} 
Here $\zeta$ is a mechanical activity parameter which is positive for extensile and negative for contractile swimmers. Active Model H thus breaks TRS via two different channels; first through the  non-equilibrium chemical potential $\mu_\A$ in~\eqref{eq:MECAMH2}, and second because, for $\zeta \neq 0$, the stress tensor and the free energy  are not related by (\ref{eq:AMH-Sigmaeq}). Our analysis covers the case where (\ref{eq:AMB},\ref{eq:MECstress1}) hold, as assumed in \cite{tiribocchi2015active}, but in fact we need no such restrictions on the form of $\cF[\phi], \mu_\A[\phi]$, and $\mathbf{\Sigma}^\A[\phi]$, so far as the latter is a symmetric tensor. 

Using the methods outlined for Active Model B in Section~\ref{sec:AMB}, we can now write the dynamical action density of Active Model H as
\begin{equation}\label{eq:AMH-action-new} 
	\al
		\cA &= \f{1}{4D} \int (\mathbb{A}_\phi+\mathbb{A}_\eta) d\bfr dt ,
		\\
	 	\mathbb{A}_\phi &= \abs{ \nabla^{-1} \brt{ (\p_t + \bv \cdot \nabla )\phi + \nabla \cdot \bfJ_d } }^2 ,
		\\
		\mathbb{A}_\eta &= \f{1}{\eta} \abs{ \nabla^{-1} \brt{ (\p_t  + {\bf v}\!\cdot\!\nabla)v_\alpha - \eta\nabla^2 v_\alpha+ \p_\alpha p - \p_\beta {\bf \Sigma}_{\alpha\beta} } }^2 
		\\
		&\quad - \f{1}{2\eta}  \Big[\nabla^{-2}\pnt{ \p_\alpha  v_\gamma  \p_\gamma v_\alpha - \p_\alpha \p_\gamma \Sigma_{\alpha\gamma}^\A + \nabla^2 p }\Big]^2\,,
	\all
\end{equation}
as shown in Appendix \ref{app:AMH-HS}.
The pressure $p$ appears in the equations of motion as a Lagrange multiplier for fluid incompressibility; accordingly $\cA$ is as given above for those trajectories $\{ \phi (\bfr, t), \bfv (\bfr, t), p (\bfr, t)\}_{0\leq t \leq \tau}$ that have $\nabla \cdot \bfv = 0$, but is infinite for all others. We next construct a time reversal transformation by supplementing \eqref{eq:MECTRS2} with
\begin{equation} \label{eq:flowreversal}
	\al
	\rho^\R (\br, t) &= \rho (\br, \tau-t) ,
	\\
	v_\alpha^\R (\br, t) &= - v_\alpha (\br, \tau-t) ,
	\\
	p^\R (\br, t) &= p (\br, \tau-t) .
	\all
\end{equation}
We note that both $\mu_\A$ and $\mathbf{\Sigma}^\A$ must be treated even under time reversal, in line with the discussion already given in Section~\ref{sec:global}.\footnote{ To understand this for Active Model H,  consider  the stress tensor defined via~\eqref{eq:MECstress1}. At first sight the above choice of time reversal appears to contradict the statement in~\cite{tiribocchi2015active} that $\zeta$ is odd under time reversal. However, this statement refers to the fact that if one reverses the direction of the fluid flow caused by a swimmer (or set of swimmers) as in~\eqref{eq:flowreversal}, this has the same effect as reversing the sign of $\zeta$, which interchanges the contractile and extensile cases. So, rather than strict time reversal, this is a conjugacy operation~\cite{seifert2012stochastic} comparing the probabilities of forward and reverse paths in two {\em different} systems.} It is then straightforward to compute the action for the time-reversed dynamics and construct the entropy production via an obvious extension of~\eqref{eq:sigma}
\begin{equation}
\label{eq:sigmav} \cS_{\phi,\bfv} = \lim_{\tau\to \infty} \cS^{\tau} ,
\quad \cS^{\tau} = \f{1}{\tau} \avg{ \log \f{ \cP[\phi,\bfv] }{
    \cP^\R[\phi,\bfv] } } , 
\end{equation}
with the following result:
\begin{equation} \label{eq:EP-AMH}
	\cS_{\phi, \bfv} = -\f{1}{D} \int \brt{ \mu_\A ( \partial_t + \bfv \cdot \nabla ) \phi + v_{\alpha\beta} \mathbf{\Sigma}^\A_{\alpha\beta} } d \bfr .
\end{equation}
Here $v_{\alpha\beta}= (\partial_{\alpha}v_{\beta}+\partial_{\beta}v_{\alpha})/2$ is the symmetrised velocity gradient tensor. This result reduces to~\eqref{eq:sigma-phi} for Active Model B when ${\bf v} = 0$, and applies under the same broad conditions. In particular, it is a steady-state quantity that excludes transient contributions. (The latter would be the only terms present in the phase ordering of passive Model H without boundary driving~\cite{Bray}.)

The Harada-Sasa relation can be further generalised to the case of Active Model H. To this end, we consider perturbations of the active parts of the chemical potential $ \mu_\A \to \mu_\A - h $ and of the stress tensor $ \mathbf{\Sigma} \to \mathbf{\Sigma} - \mathbf{\varepsilon} $, with $\mathbf \varepsilon$ symmetric and traceless. We introduce the corresponding response functions as
\begin{equation}
	\al
		\dR (\br_1,\bfr_2, t-s) &= \left. \f{ \delta \avg{ (\p_t\phi +v_\alpha \p_\alpha \phi  ) (\br_1, t) } }{ \delta h (\br_2, s) } \right|_{h=0} ,
		\\
		\cR (\br_1,\bfr_2, t-s) &= \left. \df{ \delta \avg{ v_{\alpha\beta} (\br_1, t) } }{ \delta \eps_{\alpha\beta} (\br_2, s) } \right|_{\eps=0} .
	\all
\end{equation}
We define the associated autocorrelation functions:
\begin{equation} \label{eq:AMH-corr-pt}
	\al
		\dC (\bfr_1,\bfr_2,t) &= \avg{ ( \p_t\phi + v_\alpha \p_\alpha \phi ) (\br_1, t) ( \p_t\phi + v_\alpha \p_\alpha \phi ) (\br_2,0) } ,
		\\
		\cC (\bfr_1,\bfr_2,t) &= \avg{ v_{\alpha\beta} (\bfr_1,t) v_{\alpha\beta} (\bfr_2,0) } .
	\all
\end{equation}
We demontrate in Appendix~\ref{app:AMH-HS} that the entropy production rate can be expressed as
\begin{equation} \label{eq:HS_modelH1}
	\al
		\cS_{\phi,\bfv} &= \int [\sigma_\phi({\bfk,\omega})+\sigma_{\bf v}(\bfk,\omega) ] d\bfk d\omega ,
		\\
		\sigma_\phi (\bfk, \omega) &= \f{ k^{-2} }{ D (2\pi)^{d+1} } \pnt{ \dC - 2 D \dR }(\bfk, \omega) ,
		\\
		\sigma_\bfv (\bfk, \omega) &= \f{2\eta}{ D (2\pi)^{d+1} } \pnt{ \cC - 2 D \cR } (\bfk, \omega) ,
	\all
\end{equation}
where
\begin{equation}
	\al
		\dR (\bfk, \omega) &\equiv \int \dR(\bfro,\bfrd,t) e^{i\bfk\cdot (\bfro-\bfrd)} \cos(\omega t) d \bfr_{1,2} dt ,
	\\
	\cR (\bfk, \omega) &\equiv \int \cR (\bfro,\bfrd,t) e^{i\bfk\cdot (\bfro-\bfrd)} \cos(\omega t) d\bfr_{1,2} dt .
	\all
\end{equation}

The coupling of the scalar field with an external velocity field thus requires us to consider two different FDT violations when evaluating $\cS$: one  associated with the $\phi$ dynamics, and the other with the Navier-Stokes sector. While a pair of such FDTs are satisfied independently in the passive case, the nature of the perturbations and observables entering the HSR could not have been anticipated a priori. In particular, the  chemical potential perturbation $\delta\mu = -h({\bf r},t)$ must act in the diffusive sector alone and not in the thermodynamic stress.
This requires $h$ to be considered part of $\mu_\A$ not of $\mu_\E$.
In consequence~\eqref{eq:HS_modelH1} does not connect with standard expressions for 
the (transient) entropy production in the passive limit in the way one might have expected.
This finding is relevant to the wider agenda of generalising the HSR to more complicated active field theories, for instance with orientational order~\cite{Marchetti2013RMP}: since care is required, a relatively formal approach is advisable. Our result demonstrates how the HSR framework singles out, given a set of equations of motion, exactly which FDT violations should be probed to quantify TRS breakdown.

\section{Circulating currents}
\label{sec:curl}

Active Model B, while abandoning the free energy structure of its passive counterpart, retains in \eqref{eq:general_AMB} the gradient structure of the deterministic current: $\bfJ_d = -\nabla\mu$. This implies $\nabla\wedge\bfJ_d \equiv {\bf 0}$ so that, at zeroth order in noise, there can be no circulating currents in steady state. 
 In this Section we outline how our approach generalizes to scalar field theories that allow currents of nonzero curl, as would be needed to capture (for instance) the self-assembled many-body ratched behavior reported for active Brownian particles in~\cite{stenhammar2016}. 

One such model, which will be addressed in more detail in a separate publication~\cite{AMBplus}, has $\dot\phi = -\nabla\cdot(\bfJ_d+{\bf \Lambda})$ with
\begin{eqnarray}\label{eq:BP1}
&&\bfJ_d = -\nabla\tilde\mu + [(\kappa_0+\kappa_1\phi)\nabla^2\phi+\nu|\nabla\phi|^2] \nabla\phi\\
&&{  \tilde\mu =\mu_\B+(\lambda_0+\lambda_1\phi)|\nabla\phi|^2}\nonumber
\\
&&\mu_\B=  -h+a_2\phi+a_3\phi^2+a_4\phi^3-k\nabla^2\phi .\nonumber
\end{eqnarray}
This model, which we call Active Model B+, is written here in the presence of a static free energy contribution $-h(\bfr)\phi(\bfr)$. In $\bfJ_d$ it includes, for completeness, all active terms to order $(\phi^3,\nabla^3)$. Of these, only $\kappa_1$ and $\nu$ can create nonzero curl. Indeed only $\nu$ can do this in one dimension, in the sense of creating a nonzero net current around a loop closed via periodic boundary conditions. (This is because $\phi\nabla^2\phi$ is reduces to gradient form in one dimension.) 
By Helmholtz decomposition, $\bfJ_d$ can always be written as
\begin{eqnarray}\label{eq:Jd-curly-1}
\bfJ_d &=& \bfJ_d^g+\bfJ_d^r\nonumber\\
\bfJ_d^g &= &-\nabla\mu \quad;\quad \bfJ_d^r = \nabla\wedge \bfA,
\end{eqnarray}
for some choice of scalar field $\mu$ and vector field $\bfA$, with superscripts $g,r$ standing for gradient and rotational.
By similarly decomposing the noise terms one can also formally define gradient and circulating parts of the total current $\bfJ = \bfJ^g+\bfJ^r$.

Since $\bfJ^r$ is divergence-free, it plays no part in the dynamics of $\phi$. Accordingly~\eqref{eq:sigma}, which defines $\cS_\phi$, is oblivious to any entropy production arising from circulating currents. To capture this additional source of irreversibility, we now promote $\bfJ(\bfr,t)$ into an explicit dynamical variable. This leaves $\phi$ dynamically redundant, as it can be reconstructed via $\phi(\bfr,t)= \phi(\bfr,0) - \int_0^t \nabla \cdot {\bf J}(\bfr,s) ds$. From the point of view of comparing forward and backward trajectories, this corresponds to comparing movies in which the current is recorded ({\em e.g.}, using tracer particles) rather than just density fluctuations.

If we fix a current trajectory $\{{\bf J}(x,t)\}_{0\leq t \leq \tau}$, the probability that $\bfJ$ is arbitrarily close to it has again an exponential form with action given by
\qq\label{eq:AMB-current}
\cA_\bfJ = \frac{1}{4D} \int
\left| {\bf J} - \bfJ_d\right|^2(\bfr,t) d\bfr dt.
\qqq
A straightforward computation shows that the entropy production at the level of currents is
\qq\label{eq:sigma-phi-current-rot}
\cS_\bfJ=\cS_{\phi}+ \frac{1}{D }\int \langle\bfJ^r.{\bfJ}_d^r\rangle ({\bf r})d\bfr\,.
\qqq
For comparison, note that using~\eqref{eq:variants} $\cS_\phi$ can be written  
\qq\label{eq:sigma-phi-current}
\cS_{\phi} =  \frac{1}{D }\int \langle \bfJ^g.{\bf J}_d^g\rangle({\bf r})d\bfr .
\qqq
There are no cross terms between rotational and gradient contributions because $\int \bfJ^g.(\nabla\wedge {\bf A})d\bfr$ vanishes on partial integration. (This allows either of $\bfJ^r$ or $\bfJ^g$ to be replaced by the full current $\bfJ$ in (\ref{eq:sigma-phi-current-rot},\ref{eq:sigma-phi-current}).) 
The result (\ref{eq:sigma-phi-current}) is compatible with the forms  (\ref{eq:sigma-phi},\ref{eq:sigma-phi-ito}) found above for $\sigma_\phi(\bfr)$, although $\mu^A$ is now a nonlocal functional \footnote{Its calculation involves first constructing $\mu$ as a Coulomb integral of $\nabla\cdot \bfJ$, and then identifying as $\mu_\A$ the part of it that cannot be represented as a free energy derivative.}  of $\phi$.

We listed in~\eqref{eq:variants} several, potentially inequivalent, forms for the local entropy production $\sigma_\phi$ differing by transient contributions and/or formal entropy currents. Further options arise for the correspondingly defined local quantity $\sigma_{\bfJ}$; we explore these elsewhere~\cite{AMBplus}.
Meanwhile, the simplest case of an additional entropy production arising via~\eqref {eq:sigma-phi-current-rot} is when the current contains a deterministic rotational part $\bfJ_d^r$ that is non-vanishing as $D\to 0$. The resulting contribution to $\cS_\bfJ$ is
\qq
 \frac{1}{D}\int |\bfJ_d^r|^2 d\bfr = \frac{1}{D}\int|\nabla\wedge {\bf A}|^2 d\bfr .
\qqq
This is a classical form, familiar from the case of external driving, such as a current of charged particles in a circular wire driven by a tangential electromotive force.

Interestingly,
the generalised Harada-Sasa relation (\ref{eq:HS-AMB}) for ${\cal S}_\phi$ is insensitive to the additional terms in \eqref{eq:BP1} and follows from \eqref{eq:sigma-phi} directly as before, without modification. Indeed, since the circulating currents have no effect on the dynamics of $\phi$, the FDT mismatch of its correlators can give no information about their contribution to the entropy production via the last term in (\ref{eq:sigma-phi-current-rot}). A connection between Harada-Sasa theory and this term might be made, by considering the fluctuations of the curly currents themselves and their response to perturbations that couple to them. We leave this for future investigations.

\section{Conclusions}

\label{Conc}

We have studied the violation of time reversal symmetry (TRS) in a class of scalar field theories relevant for the description of active matter such as self-propelled particles without alignment interactions. We have presented general tools for addressing the question of {\it  whether activity matters dynamically}---meaning, whether it contributes distinctly to TRS-breaking physics, or whether the dynamics at large scales could be reproduced by some (possibly complicated) equilibrium model with TRS. Studying the entropy production allows one to quantify TRS breakdown directly at coarse-grained level: it unambiguously assesses the extent to which dynamics at this scale is genuinely nonequilibrium. 

The main models we considered were Active Models B and H~\cite{wittkowski2014scalar, tiribocchi2015active}. For Active Model B, we defined a local entropy production density and confirmed this to be strongly localized at the interface between phases, with a bulk contribution that is much weaker at low noise levels, by constructing spatial maps to quantify \emph{where} activity plays a role. We also offered a generalisation of the Harada-Sasa relation (HSR) to field theories, quantitatively relating the entropy production rate to the violation of the fluctuation-dissipation theorem (FDT). This allows entropy production to be spectrally decomposed across spatial and temporal Fourier modes, even in translationally invariant cases where real-space maps convey no useful information.

Assigning wave-number dependence to an entropy production allows one to give quantitative underpinning to a statement like ``this active system is effectively passive, but only at scales larger than the correlation length''. The quantitative link between FDT violation and TRS breakdown reinforces the fact that TRS is, first, a scale-dependent phenomenon, and second, one that depends on which variables are retained in a coarse-grained description. Absence of FDT violation ensures zero entropy production, but only if the FDT is tested across the full subset of dynamical fields that we wish to describe. As we showed for Active Model H, the requirements are not always obvious. It might therefore be wise to focus first on the entropy production, and only when the form of this is clear turn to Harada-Sasa type constructions to make an explicit connection to FDT violations. This should ensure that no important terms are overlooked. An open task is to construct a HSR capable of detecting the entropy production from circulating currents that are invisible in the dynamics of the density field $\phi(\bfr, t)$, as exemplified by Active Model B+.

In scalar active matter  the micro-dynamics is usually very far from equilibrium, despite which several approximate mappings to equilibrium. Coarse-grained approximations for the dynamics have been proposed in a number of systems \cite{bertin2006boltzmann,liverpool2005bridging,bertin2013mesoscopic,ihle2011kinetic}, establishing at times a surprising connection to equilibrium \cite{Tailleur:08,thompson2011lattice}. Note the question of whether a given non-TRS dynamics can be approximated, at coarse grained level, by one with TRS is a distinct issue from whether the original microscopic dynamics is close to equilibrium \cite{bo2014entropy}. In this article we provide the tools to decided whether TRS breaking survives coarse-graining in active matter systems.

In some active matter systems, mesoscopic or macroscopic violations of TRS are obvious -- for instance when fluid vortices are visible at the scale of interest. Our machinery can quantify such violations, but is likely to be more useful in cases where they are harder to detect. For example, it would be an interesting follow up of this work to apply the theoretical results developed here to Model B coupled to a logistic birth/death term which causes arrest of phase separation at a particular scale~\cite{CatesPNAS}. Particles (representing bacteria) divide in dilute regions and die off in dense ones, manifestly violating TRS at the level of currents. It would be interesting to know if such a model still has finite entropy production when the density field alone is monitored.

Finally, there is clearly a connection between the idea of scale-dependent entropy production and the renormalization-group concept of ``relevance''. However, these are not interchangeable, since activity could be relevant (changing a universality class) without itself surviving coarse-graining (the new class might restore TRS at its fixed point). We hope our work will inform future RG studies both of scalar active field theories and the more complex vector and tensor theories of active matter~\cite{Marchetti2013RMP, ChenTonerLee2015, mishra2010dynamic, toner1995long, chen2013universality}, for which RG studies of are starting to systematically uncover new classes.


\acknowledgments

We thank Hugues Chat\'e, Romain Mari, Fred MacKintosh, Davide Marenduzzo, Thomas Speck, David Tong, Paolo Visco and Raphael Wittkowski for illuminating discussions. C. Nardini acknowledges the hospitality provided by DAMTP, University of Cambridge while this work was being done. This research has been supported by EPSRC grant EP/J007404. MEC holds a Royal Society Research Professorship.


\appendix

\section{Additivity and locality of entropy production}

\label{app:sec:local}

For Active Model B and its relatives we established relations such as (\ref{eq:Sdensity},\ref{eq:mec1a}) relating the global steady state entropy production rate $\cS$ to the integral of a local quantity $\sigma(\bfr)$. 
This shows that the entropy production is additive over regions of space within a larger system. However, if we were to isolate one of these regions as a subsystem, its entropy production rate would depend on what boundary conditions were imposed. Only in the thermodynamic limit where all subsystems are much larger than the correlation length, do the contributions from such internal boundaries become unimportant. 

In this Appendix we consider the case where one specifies fixed information about the time evolution of $\phi$ and its gradients on each internal boundary which is then shared by the subsystems on either side of that boundary. We examine the additivity of the entropy production in this case, and discuss conditions under which it makes sense to view $\sigma(\bfr)$ as a local entropy production density.

We consider a region of space $\Omega_1$, with $\Omega_2$ its complement; $\p \Omega$ is the boundary separating the two regions. We restrict attention to local field theories, defined as those in which the trajectories  within $\Omega_1$ have no further dependence on those in $\Omega_2$ once the boundary information $\bchi$ is given. The probability for the system to follow a given trajectory $\bpsi$ which equals $\bpsi_1(\bfr,t)$ inside $\Omega_1$, and $\bpsi_2(\bfr,t)$ in $\Omega_2$, can then be written as  
\begin{equation}\label{eq:AMB-PJ-conditional}
\mathcal{P}[\bpsi] = \mathcal{P}_1[\bpsi_1|\bchi]\,\mathcal{P}_2[\bpsi_2|\bchi]\,\mathcal{P}_{\p\Omega}[\bchi]\,,
\end{equation}
where $\mathcal{P}_i[\bpsi_i|\bchi]$ is the conditional probability of finding $\bpsi_i$ in $\Omega_i$, given that these fields and their derivatives satisfy some boundary information $\bchi(t)$ on $\p \Omega$, whose unconditional probability is $\mathcal{P}_{\p\Omega}[\bchi]$.

 For Active Model B, Eq.~\eqref{eq:AMB-PJ-conditional}  is obviously true if one works with the current ($\bpsi = {\bf J}$) of which the action $\cA[\bfJ]$ is a local functional. Less obviously, this is true also working with the density $\phi$, so long as the boundary data $\bchi$ contains enough derivatives of $\phi$ both to specify $(\dot \phi + \nabla \cdot {\bf J}_d)$ on $\p\Omega$ and to ensure that the Poisson equation $\nabla^2 f = (\dot \phi + \nabla \cdot {\bf J}_d)$ has an unique solution inside $\Omega_i$. In such cases one has
\begin{equation}\label{eq:weightmec}
\mathcal{P}_i[\bpsi_i|\bchi] = \frac{W_i[\bpsi_i|\bchi]}{Z_i[\bchi]}
\end{equation}
where $W_i$ is a statistical weight for the conditioned trajectories in $\Omega_i$, and $Z_i[\bchi]$ its integral over such trajectories. For any local action $\cA$, we have $W_i = \exp-\cA_i[\bpsi_i]$ for trajectories respecting the boundary data and $W_i = 0$ otherwise; here $\cA_i$  is an action integrated only over the domain $\Omega_i$. From this follows the statistical weight of the boundary data itself, $W_{\p\Omega}[\bchi] = Z_1[\bchi]Z_2[\bchi]$, and its normalized probability density
\begin{equation}\label{eq:Pchi}
\mathcal{P}_{\p\Omega}[\bchi] = \frac{Z_1[\bchi] Z_2[\bchi]}{\int Z_1[\bchi]Z_2 [\bchi]\;{\cal D} [\bchi]}.
\end{equation}
Substituting (\ref{eq:weightmec},\ref{eq:Pchi}) in~\eqref{eq:AMB-PJ-conditional} for the forward probability $\cP[\bpsi]$, and then repeating exactly the same calculation for the backward probability $\cP^\R[\bpsi]$, gives 
\begin{equation}\label{eq:AMB-EP-splitting}
\cS = 
\left\langle \cS_1[\bchi]  \right\rangle_{\mathcal{P}_{\p\Omega}}
+\left\langle \cS_2[\bchi]  \right\rangle_{\mathcal{P}_{\p\Omega}}
+\cS_\bchi
\end{equation}
where the entropy production of the boundary process is written
\begin{equation}
\cS_\bchi
=\lim_{\tau\to\infty}\frac{1}{\tau}\left\langle \log\frac{\mathcal{P}_{\p\Omega}}{\mathcal{P}_{\p\Omega}^\R}  \right\rangle_{\mathcal{P}_{\p\Omega}}.
\end{equation}
In \eqref{eq:AMB-EP-splitting}, the first two terms sum the entropy productions in our two subsystems; each of these is first calculated with quenched boundary data $\bchi$ as
\begin{equation}
\cS_i[\bchi] =\lim_{\tau\to\infty} \cS_i^{\tau}[\bchi]\qquad
\cS_i^{\tau}[\bchi]=\frac{1}{\tau}\left\langle \log \frac{\mathcal{P}_i}{\mathcal{P}_i^\R} \right\rangle_{\mathcal{P}_i[\bpsi_i|\bchi]}
\end{equation}
and then averaged over $\bchi$ (notated as $\langle\cdot\rangle_{\mathcal{P}_{\p\Omega}}$ in~\eqref{eq:AMB-EP-splitting}). 

From~\eqref{eq:weightmec} we obtain 
\begin{equation}\label{eq:deco}
	\al
		&\cS_i[\bchi] =
		\\
		& = \left\langle\int_{\Omega_i} \hat\sigma(\bpsi,\nabla\bpsi,...) d\bfr  \right\rangle_{\mathcal{P}_i[\bpsi_i|\bchi]} +\lim_{\tau\to\infty}\frac{1}{\tau}\log\frac{Z_i^\R[\bchi]}{Z_i[\bchi]}
	\all
\end{equation}
with $\hat\sigma(\bpsi,\nabla\bpsi,...)$ a local function of $\bpsi$ and its gradients. The first term represents the bulk contribution to $\cS$ while the second only depends on the boundary data. This does not vanish in general, but it does so whenever the boundary information $\bchi$ itself exhibits TRS. 

Thus the entropy production rate $\cS$ for two systems with a shared boundary is strictly additive only if a movie of events on the boundary itself looks statistically the same when shown in reverse. Generically this cannot be expected, so that there is indeed a finite (sub-extensive) boundary contribution, which becomes negligible only if the subsystems are large compared to the correlation length. 

This type of `qualified additivity' is familiar from the study of systems in equilibrium. Indeed, using arguments that exactly parallel those above, one can show that in equilibrium the global entropy defined as
\begin{equation}\label{gibbsent}
S = -\int \mathcal{P}[\bpsi]\ln(\mathcal{P}[\bpsi]){\cal D}[\bpsi]
\end{equation}
obeys (analogous to \eqref{eq:AMB-EP-splitting})
\begin{equation}\label{eq:AMB-EP-splitting_eqm}
S = 
\left\langle S_1[\bchi]  \right\rangle_{\mathcal{P}_{\p\Omega}}
+\left\langle S_2[\bchi]  \right\rangle_{\mathcal{P}_{\p\Omega}}
+ S_\bchi
\end{equation}
with the boundary contribution to the entropy
\begin{equation}
S_\bchi = - \int \mathcal{P}_{\p\Omega}[\bchi]\ln(\mathcal{P}_{\p\Omega}[\bchi])\,{\cal D}[\bchi]\,.
\end{equation}
Moreover, in analogy with \eqref{eq:deco} we find
\begin{equation}\label{eq:deco_eqm}
		S_i[\bchi] =
		 -\left\langle\ln W[\bpsi_i |\bchi] \right\rangle_{\mathcal{P}_i[\bpsi_i|\bchi]} + \ln Z_i[\bchi]
\end{equation}
In (\ref{gibbsent}-\ref{eq:deco_eqm}), $S_i, \mathcal{P}$ and $W$ are now functionals of configurations and not trajectories.

These arguments for qualified additivity support the interpretation of $\sigma(\bfr) = \langle\hat\sigma\rangle$ in~\eqref{eq:Sdensity} as a local entropy density, on the basis that $\hat\sigma(\bpsi,\nabla\bpsi,...)$ stands conceptually in relation to the global entropy production $\cS$ just as  the equilibrium entropy density ${\mathbb S}(\bpsi,\nabla\bpsi,...)$ stands in relation to  the global equilibrium entropy $S$. Note that in the main text we have mainly focused on the weak noise limit, for which the integral in~\eqref{eq:Sdensity} is performed at a sufficiently coarse-grained level that long-wavelength fluctuation contributions are unimportant. In equilibrium models this would correspond to a treatment of $\mathbb S(\bpsi,\nabla\bpsi,...)$ at mean-field or density-functional level.

Having established qualified additivity, and given the non-negativity of $\cS$, a natural question is whether the entropy production density $\sigma(\bfr)$ is itself non-negative. In partial answer to this, observe first that the various contributions in~\eqref{eq:AMB-EP-splitting} are non-negative since the relevant integral fluctuation theorems~\cite{seifert2012stochastic} apply equally to conditional and full probabilities. We thus have:
\begin{equation}
	\cS_i^{\tau} [\bchi] \geq0 , \quad \f{1}{\tau} \avg{ \log \f{\cP_{\p\Omega} }{ \cP_{\p\Omega}^\R } }_{ \cP_{\p\Omega} } \geq0 .
\end{equation}
This might seems to imply the positivity of $\sigma(\bpsi,\nabla\bpsi,...)$ when the continuum limit is taken. However, due to the presence of the second term of~\eqref{eq:deco}, this is actually not a correct inference unless the statistics of the boundary information $\bchi$ itself exhibits TRS, so that the second term vanishes. This suggests (as is physically reasonable) that any local region of negative $\sigma$ must be driven, through its boundary, by a larger, positive entropy production happening elsewhere. We defer to future work the question of whether such situations can arise in practice for the active field theories studied here.


\section{Discretised dynamics}

\label{app:discretised}

In this Appendix, we detail how to perform spatial discretisation such that detailed balance is always recovered in the equilibrium limit for Active Model B. This is the discretisation we used to numerically integrate the model, giving the results in Sec.~\ref{sec:AMB}.

We consider here 1D Model B ($\mu_\A=0$), although it is easy to extend the results of this Appendix to higher dimensions. We consider a system of finite width $L$ such that $x\in[0,L]$ with periodic boundary conditions. We discretise $x$ into $N$ lattice points with equal lattice spacing $\Delta $ so that $N\Delta =L$, and the density field as $\phi(x,t)\rightarrow\phi_{i}(t)$, where $i=1,2,\ldots N$; $\phi_{i}$ is the value of $\phi$ at $x=i\Delta $. The dynamics then becomes:
\begin{equation}\label{eq:discrete1}
	\p_t \phi_i = \nabla^2 \f{ \p \cF }{ \p \phi_i} + \sqrt{2D} \nabla \eta_i ,
\end{equation}
with $ \avg{ \eta_i(t) \eta_j (t') } = \delta_{ij} \delta(t-t') $. We now choose the following discretisation for the gradient operator $\nabla$:
\begin{equation}\label{eq:grad}
	\nabla \psi_i = \f{ \psi_{i+1} - \psi_{i-1} }{ 2 \Delta } ,
\end{equation}
for any discrete fields $\psi_{i}$, which implies the discretisation for the Laplacian operator:
\begin{equation}\label{eq:laplacian}
	\nabla^{2}\psi_{i}=\frac{\psi_{i+2}-2\psi_{i}+\psi_{i-2}}{4\Delta ^{2}} .
\end{equation}
To show detailed balance is satisfied, we substitute Eqs.~(\ref{eq:grad},\ref{eq:laplacian})
into~\eqref{eq:discrete1} to obtain:
\begin{equation}
	\p_t \phi_i = - \f{1}{ 4 \Delta^2 } A_{ij} \f{ \p \cF }{ \p \phi_j} + \f{\sqrt{2D}}{2\Delta} B_{ij} \eta_j ,
\end{equation}
where the matrices $A$ and $B$ are given by:
\begin{equation}
A = \pnt{ \begin{array}{cccccc}
2 & 0 & -1\\
0 & 2 & 0 & -1\\
-1 & 0 & 2 & 0 & -1\\
 &  &  & \ddots\\
 &  &  &  & \ddots\\
 &  &  & -1 & 0 & 2
\end{array} } ,
\end{equation}
and 
\begin{equation}\label{app:discr-B}
B = \pnt{\begin{array}{ccccc}
0 & 1\\
-1 & 0 & 1\\
 & -1 & 0 & -1\\
 &  &  & \ddots\\
 &  &  & -1 & 0
\end{array}} ,
\end{equation}
with zero elements where not explicitly written. We observe $BB^{T}=B^{T}B=A$, which is the condition for this discretised dynamics to obey exactly detailed balance~\cite{Gardiner:1985,VanKampenBook}. Using the same discretization for the active term $\mu_\A$, this ensures that TRS is respected fully in the limit $\mu_\A\to 0$.


\section{Time-symmetric contribution to ${\cal A}$}

\label{app:discretised-strato-action}

The time symmetric contribution that was omitted from~\eqref{eq:intro-S} depends on the prescription used for defining stochastic integrals (Ito, Stratonovich, or intermediate \cite{Lau:07,aron2014dynamical,aron2010symmetries,janssen1979}) and also on the spatial discretisation used; indeed subscript $S$ denotes our adopted Stratonovich convention.  As discussed in Appendix~\ref{app:discretised} above, we use midpoint discretization to maintain exact TRS in the passive limit. We therefore present here only the form of this term, notated $\cA_S$ in the following, for Active Model B in the Stratonovich convention with midpoint discretization and the choice $\mu_\A=\lambda |\nabla \phi|^2$. This is straightforwardly obtained from Eq. (5.15) in~\cite{Lau:07} as 
\begin{equation} \label{eq:app-AS}
	\cA_S= \cA_S^\R = - \f{1}{2\Delta^2} \sum_i \brt{ f''(\phi_i) + \f{3\kappa}{4\Delta^2} } .
\end{equation}
This is divergent as $\Delta \to 0$, both in the part that depends on $f''(\phi_i)$ and in the second, constant contribution. However one still has $ \cA_S - \cA_S^\R = 0$, as promised, which is all we need for the results of the main text. The use of any other convention and/or discretization would give the same final result for $\cS$ once the counterpart of $ \cA_S - \cA_S^\R$, which is no longer zero, is properly worked out.


\section{Spatial discretization}

\label{app:discretised-ito-EP}

The result~\eqref{eq:sigma-phi-ito} follows from~\eqref{eq:sigma-phi} only if $\langle  \mu_\A\nabla.{\bf \Lambda}\rangle = {0}$. This would hold automatically if the stochastic integrals were interpreted in the Ito convention in which the time derivative is evaluated at the start of the timestep; however in this paper we use the (mid-timestep) Stratonovich convention, which anticipates part of the subsequent increment so that $\dot\phi$ cannot be simply replaced by $\nabla^2 \mu$ to give~\eqref{eq:sigma-phi-ito} from \eqref{eq:sigma-phi}. Here we address only the special case of Active Model B in one dimension. We show that~\eqref{eq:sigma-phi-ito} still holds, so long as we employ mid-point spatial discretisation.

Any stochastic integral in the Stratonovich convention can be transformed in one in the Ito convention. Subtleties however arise when dealing with stochastic PDEs; these are closely linked to the $\cA_S$ term in the dynamical action discussed in Appendix~\ref{app:discretised-strato-action} above. Let us first consider a stochastic differential equation in the form ($x_i\in \mathbb{R}, x=(x_1,...,x_n)$)
\begin{equation} \label{eq:sde-1}
	\dot{x}_i = a_i(x) + b_{ij} \eta_j ,
\end{equation}
where $ \avg{ \eta_i(t) \eta_j(s) } = \delta_{ij} \delta(t-s)$. We want to consider the following Stratonovich integral
\begin{equation}
	\mathcal{I}_{il} =  \int_0^t f_i(x(s)) \eta_l(s) d s,
\end{equation}
where $x(s)$ satisfies~\eqref{eq:sde-1}. $\mathcal{I}_{il}$ can be converted in an Ito plus a non-stochastic (Riemann) integral as follows~\cite{kloeden2012numerical}
\begin{equation}\label{eq:Ito-Strato-finite-d}
	\mathcal{I}_{il} =  \int_0^t f_i(x(s)) \cdot \eta_l(s) ds + \mathcal{I}_{il}^{\textrm{conv}} ,
\end{equation}
where $\cdot$ denotes that the integral has to be understood in the Ito sense, and
\begin{equation}	\label{eq:app-Ito-Strato-conv}
	\mathcal{I}_{il}^{\textrm{conv}} = \f{1}{2} \int_0^t \f{\partial f_i}{\partial x_j}(x(s)) b_{jl} ds .
\end{equation}
This result, when formally generalised to the case of Active Model B, produces ill-defined formulae, involving the square of a Dirac delta. In order to progress, we then midpoint-discretise the dynamics (\ref{eq:general_AMB})-(\ref{eq:AMB}) as in Appendix~\ref{app:discretised}, with periodic boundary conditions. We have
\begin{equation}
	\dot{\phi}_i = \f{ \mu_{i+2}-2\mu_i+ \mu_{i-2} }{4\Delta^2} + \f{\sqrt{2D}}{2\Delta} B_{ij}\eta_j ,
\end{equation}
where $B$ obeys~\eqref{app:discr-B}, that is $B_{ij}= \delta_{i+1,j} - \delta_{i-1,j}$. Now consider the part of the entropy production coming from the stochastic integral:
\begin{equation}
\mathcal{I}=-\frac{\Delta}{2tD \Delta}\sum_i \int_{0}^t \mu_{\A,i} (\eta_{i+1}-\eta_{i-1}) .
\end{equation}
Applying (\ref{eq:app-Ito-Strato-conv}) and using the discrete form of $\mu_\A$
\begin{equation}\label{eq:mu-i}
	\mu_{\A,i} = \lambda \pnt{ \f{ \phi_{i+1} - \phi_{i-1} }{2\Delta} }^2 ,
\end{equation}
we have $ \mathcal{I}^{\textrm{conv}} = 0 $. We conclude that, with midpoint spatial discretisation, the expression for the entropy production~\eqref{eq:sigma-phi} can be equally interpreted in Ito or Stratonovich conventions. Then, using the non-anticipating property of the Ito convention,~\eqref{eq:sigma-phi} implies~\eqref{eq:sigma-phi-ito}.

This result requires the integrand $\mu_{\A,i}$ in~\eqref{eq:sigma-phi} to depend only on $i\pm 1$ and not on $i\pm 2$. If we wanted to write the entropy production in the equivalent form
\begin{equation} \label{eq:sigma-phi-app}
	\cS = - \f{1}{D} \int \langle \mu \dot \phi \rangle ({\bfr}) d\bfr ,
\end{equation}
we would find (with midpoint spatial discretisation) that the conversion from Stratonovich to Ito brings in a term
\begin{equation} \label{eq:sigma-phi-app-2}
	- \lim_{t\to\infty} \frac{1}{t} \frac{\Delta}{2\Delta^2} \sum_i \int \brt{ \frac{3 \kappa}{4\Delta^2} +  f''(\phi_i) } dt ,
\end{equation}
which does not admit a continuous limit for $\Delta\to0$. However this divergence cancels against another term
\begin{equation}
 - \f{1}{D} \int \langle \mu \cdot \nabla \Lambda  \rangle ({\bfr}) d \bfr ,
\end{equation}
that appears in the Ito but not the Stratonovich integral for $\cS$. These fact have been checked in our numerical simulations of Active Model B. Let us finally observe that~\eqref{eq:sigma-phi-app-2} is actually proportional to $\cA_S$; see~\eqref{eq:app-AS}. This is however non generic and due to the fact that the noise in the dynamics is additive and $BB^{T}=B^{T}B=A$.

We conclude by noting that analogous apparent divergences would be found using other type of spatial discretisation or Fourier truncations in the numerical computation of the entropy production.


\section{Small noise expansion}

\label{app:small_noise}
We expand the dynamic action with the $D$-expansion of $\phi$
in~\eqref{eq:AMB-phi-decomposition}. Since we use this action only for bookkeeping, we choose the simpler
It\=o convention in this appendix. We have 
\begin{equation}\label{eq:action_expanded}
 \al \cA[\phi] &= -\f{1}{4D} \int
 \brt{ \dot \phi_0 + \nabla \cdot \bfJ_d [\phi_0] } \\
&\qquad\quad  \nabla^{-2}
  \brt{ \dot \phi_0 + \nabla \cdot \bfJ_d [\phi_0] }
 d \bfr dt \\
&-
 \f{1}{2\sqrt{D}} \int
 \brt{ \dot \phi_0 + \nabla \cdot \bfJ_d [\phi_0] } 
 \nabla^{-2} \\
 &\qquad\qquad\quad
  \brt{ \dot \phi_1 - \nabla^2 \pnt{ \f{ \delta \cF_0 }{ \delta \phi_1 } + 2 \lambda \nabla \phi_0 \cdot \nabla \phi_1 } }
 d \bfr dt\\
 &-
 \f{1}{4} \int
\brt{ \dot \phi_1 - \nabla^2 \pnt{ \f{ \delta \cF_0 }{ \delta \phi_1 } + 2 \lambda \nabla \phi_0 \cdot \nabla \phi_1 } }
 \nabla^{-2} \\
  &\qquad\quad
\brt{ \dot \phi_1 - \nabla^2 \pnt{ \f{ \delta \cF_0 }{ \delta \phi_1 } + 2 \lambda \nabla \phi_0 \cdot \nabla \phi_1 } }  d\bfr dt\\
&+\cO (\sqrt{D})\,,
 \all
\end{equation}
where we have used the definition of $\cF_0$ in~\eqref{eq:F_0}. In the small noise limit, the first term in~\eqref{eq:action_expanded} must vanish to avoid any divergence, yielding the mean-field equation: $ \dot \phi_0 = - \nabla \cdot \bfJ_d (\phi_0) $. The second term thus also vanishes and the third one 
corresponds to the dynamics of $\phi_1$ given by~\eqref{eq:dynphi1}.


\section{Harada-Sasa for Active Model H}

\label{app:AMH-HS}
We first prove that the action of Active Model H is given by (\ref{eq:AMH-action-new}). Introducing the noise vector $ \Upsilon_\alpha = \p_\beta \Gamma_{\alpha\beta} $, its correlations read
\begin{equation}
\begin{aligned}
	&\avg{ \Upsilon_\alpha (\br, t) \Upsilon_\beta (\br', t') } =\\
	&= - 2 \eta D \pnt{ \delta_{\alpha\beta} \nabla^2 + \p_\alpha \p_\beta } \delta (\br-\br') \delta (t-t') \,.
\end{aligned}
\end{equation}
 The dynamic action corresponding to the fluid dynamics is
\begin{equation}
	\cA_{\eta} = \f{1}{2} \int_{\br, \br', t, t'} \Upsilon_\alpha (\br, t) \Xi_{\alpha\beta} (\br-\br', t-t') \Upsilon_{\beta} (\br', t') ,
\end{equation}
where
\begin{equation}
\begin{aligned}
	&\int_{\br'', t''} \Xi_{\alpha\gamma} ( \br-\br'', t-t'' ) \avg{ \Upsilon_{\gamma} (\br'', t'') \Upsilon_{\beta} (\br', t') } = \\
	&=\delta_{\alpha\beta} \delta (\br-\br') \delta (t-t') .
\end{aligned}
\end{equation}
We get the explicit expression for $\Xi$ as
\begin{equation}
\begin{aligned}
	&\Xi_{\alpha\beta} ( \br-\br', t-t' ) =\\
	&= - \f{1}{2 \eta D} \nabla^{-2} \pnt{ \delta_{\alpha\beta} - \f{1}{2} \nabla^{-2} \p_\alpha \p_\beta } \delta (\br-\br') \delta (t-t') \,.
\end{aligned}
\end{equation}
Using incompressibility, the contribution of the fluid dynamics to the total action follows as in (\ref{eq:AMH-action-new}).

We now consider the derivation of the generalised Harada-Sasa relation for Active Model H, stated in~\eqref{eq:HS_modelH1}. It is useful to start from an expression of the entropy production equivalent to~\eqref{eq:EP-AMH} up to boundary terms:
\begin{equation}	\label{eq:EP-AMH-app}
	\cS_{\phi, \bfv} = - \f{1}{D} \int \avg{ \mu (\p_t + \bfv \cdot \nabla ) \phi + v_{\alpha\beta}\mathbf{\Sigma}_{\alpha\beta} } d \bfr .
\end{equation}
This can be rewritten as
\begin{eqnarray} \label{eq:sigma_H}
\cS_{\phi,{\bf v}} &=& - \df{1}{2D} \underset{t\to0}{\lim} \int
\Big\langle \mu (\br, 0) \brt{ \pnt{ \p_t + v_\alpha \p_\alpha} \phi } (\br, t)\nonumber \\
&+&   \mu (\br, t) \brt{ \pnt{ \p_t + v_\alpha \p_\alpha} \phi } (\br, 0) 
\\
&+ &   v_{\alpha\beta} (\br, 0) \mathbf{\Sigma}_{\alpha\beta} (\br, t) + v_{\alpha\beta} (\br, t) \mathbf{\Sigma}_{\alpha\beta} (\br, 0) \nonumber\Big\rangle d\bfr \nonumber
\end{eqnarray}
The dynamics perturbed as $ \mu_\A \to \mu_\A - h $ is given by
\begin{equation}
	\pnt{ \p_t + v_\beta \p_\beta } \phi = - \p_\alpha J_\alpha , \quad J_\alpha = - \p_\alpha \pnt{ \mu - h } + \Gamma_\alpha .
\end{equation}
The corresponding dynamical action shift at linear order in $h$ reads
\begin{equation}
	\delta \cA = - \f{1}{2D} \int h \brt{ \pnt{ \p_t + v_\beta \p_\beta } \phi - \p^2_{\alpha\alpha} \mu } d\br dt + \cO \pnt{ h^2 } .
\end{equation}
From this follows the response
\begin{equation}
	\al
		&\dR  (\br,\br, t) =
		\\
		& \f{1}{2D} \avg{ \brt{ \pnt{ \p_t + v_\alpha \p_\alpha } \phi } (\br, t) \brt{ \pnt{ \p_t + v_\beta \p_\beta } \phi - \p^2_{\beta\beta} \mu } (\br, 0) } ,
	\all
\end{equation}
and its symmetrized form
\begin{equation} \label{eq:resp_H1}
	\al
		&\dR(\br,\br, t)+ \dR(\br,\br, -t) =
		\\
		& \f{1}{D} \avg{ \brt{ \pnt{ \p_t + v_\alpha \p_\alpha } \phi } (\br, t) \brt{ \pnt{ \p_t + v_\beta \p_\beta } \phi } (\br, 0) }
		\\
		& - \f{1}{2D} \nabla^2_{(2)} \avg{ \brt{ \pnt{ \p_t + v_\alpha \p_\alpha } \phi } (\br, t) \mu (\br, 0) }
		\\
		& -\f{1}{2D} \nabla^2_{(2)} \avg{ \brt{ \pnt{ \p_t + v_\alpha \p_\alpha } \phi } (\br, 0) \mu (\br, t) } .
	\all
\end{equation}
We see that the first two lines of~\eqref{eq:sigma_H} can be written
\begin{equation}
	\al
		& \f{1}{2} \avg{ \brt{ \pnt{ \p_t + v_\alpha \p_\alpha } \phi } (\br, t) \mu (\br, 0) } +
		\\
		& \f{1}{2} \avg{ \brt{ \pnt{ \p_t + v_\alpha \p_\alpha } \phi } (\br, 0) \mu (\br, t) }
		\\
		& = \nabla^{-2}_{(2)} \dC (\br,\br, t) - D\nabla^{-2}_{(2)} \brt{ \dR (\br,\br, t) + \dR (\br,\br, -t) } ,
	\all
\end{equation}
where we employed the definition~\eqref{eq:AMH-corr-pt} for the correlation.

To obtain the last line in the expression for the entropy production~\eqref{eq:sigma_H}, we now consider that the response with respect to the perturbation $ {\bf \Sigma}^\A \to {\bf \Sigma}^\A - \mathbb{ \eps} $ with $\eps_{\alpha\beta}$ symmetric in the exchange of $\alpha$ and $\beta$ and traceless. The dynamic action at linear order in $\eps$ reads 
\begin{equation}\label{eq:AMH-app-delta-A}
	\begin{aligned}
		\delta \cA =&- \f{1}{2 \eta D} \int_{\br, t} \p_\beta \eps_{\alpha\beta} \nabla^{-2} \brt{  \pnt{ \p_t + v_\gamma \p_\gamma } v_\alpha \right.\\
&\qquad\qquad\left.- \eta \nabla^2 v_\alpha - \p_\gamma \Sigma_{\alpha\gamma}^\A + \p_\alpha p }
		\\
		& - \f{1}{4 \eta D} \int_{\br, t} \p_\alpha \p_\gamma \eps_{\alpha\gamma} \nabla^{-4} \pnt{ \p_\beta   v_\mu  \p_\mu v_\beta \right.\\
		&\qquad\qquad\left.- \p_\beta \p_\mu \Sigma_{\beta\mu}^\A + \nabla^2 p } + \cO (\eps^2) \,.
	\end{aligned}
\end{equation}
Now employing a formula analogous to~\eqref{eq:AMB-response-implicit}, the response follows as
\begin{equation}\label{eq:app-HS-AMH-hysata}
	\al
		 \cR (\br, \br, t) =& - \f{1}{2 \eta D} \avg{ v_{\alpha\beta} (\br, t)\, \p_\beta \nabla^{-2} \brt{  \pnt{ \p_t + v_\gamma \p_\gamma } v_\alpha \right.\right.\\
		&\qquad\qquad\left.\left.- \eta \nabla^2 v_\alpha - \p_\gamma \Sigma_{\alpha\gamma}^\A + \p_\alpha p } (\br, 0) }
		\\
		& + \f{1}{4 \eta D} \avg{ v_{\alpha\beta} (\br, t)\, \p_\alpha \p_\beta \nabla^{-4} \pnt{ \p_\gamma   v_\mu  \p_\mu v_\gamma \right.\right.\\
		&\qquad\qquad\left.\left.- \p_\gamma \p_\mu \Sigma_{\gamma\mu}^\A + \nabla^2 p } (\br, 0) } .
	\all
\end{equation}
We now observe that we only need the integral over space:
\begin{equation}
	\al
		& \int \cR (\br, \br, t) d \bfr = \f{1}{2D} \int \avg{ v_{\alpha\beta}(\br, t) v_{\alpha\beta}(\br, 0) } d \bfr 
		\\
		& \qquad+ \f{1}{4 \eta D} \int \avg{ v_{\alpha\beta}(\br, t) {\bf \Sigma}_{\alpha\beta} (\br, 0) } d \bfr
		\\
		& \qquad+ \f{1}{4 \eta D} \int \avg{ v_\alpha (\br, t)  \pnt{ \p_t + v_\gamma \p_\gamma } v_\alpha  (\br, 0) } d \bfr ,
	\all
\end{equation}
where we have used the incompressibility condition and integration by parts to eliminate both the pressure term and the contribution given by the third and the fourth line of (\ref{eq:app-HS-AMH-hysata}). In the $t\to0$ limit, we have
\begin{equation}
	\al
		& \lim_{t\to0} \int \avg{ v_\alpha (\br, t)  \pnt{ \p_t + v_\gamma \p_\gamma } v_\alpha  (\br, 0) } d \bfr=
		\\
		&\quad =  \int \avg{ \f{1}{2} \p_t v^2_\alpha (\br, 0) + v_\alpha v_\gamma \p_\gamma v_\alpha (\br, 0) } d \bfr = 0 .
	\all
\end{equation}
Using the definition of $\dC$ in~\eqref{eq:AMH-corr-pt} we obtain
\begin{equation} \label{eq:resp_H2}
	\al
		\lim_{t\to0} \int & \brt{ \cR (\br, \br, t) + \cR (\br, \br, -t) } d \br = \f{1}{D} \lim_{t\to0} \int \dC (\br, \br, t) d\br
		\\
		& + \f{1}{4\eta D} \lim_{t\to0}\ \int \avg{ v_{\alpha\beta}(\br, t) {\bf \Sigma}_{\alpha\beta}(\br, 0) } d\br
		\\
	  & + \f{1}{4\eta D} \lim_{t\to0}\ \int \avg{ v_{\alpha\beta} (\br, 0) {\bf \Sigma}_{\alpha\beta}(\br, t) } d \br .
	\all
\end{equation}
Plugging Eqs.~\eqref{eq:resp_H1} and~\eqref{eq:resp_H2} in~\eqref{eq:sigma_H}, we finally get
\begin{equation} \label{eq:AMH-app-final}
	\al
		& \cS_{\phi,\bfv} =
		\\
		& \f{1}{D} \underset{t\to0}{\lim} \int \nabla^{-2}_{(2)} \cur{ D \brt{ \dR (\br, \br, t) +  \dR (\br, \br, -t) } - \dC (\br, \br, t) } d \br
		\\
		& + \f{2\eta}{D} \underset{t\to0}{\lim} \int \cur{ \cC (\br, \br, t) - D \brt{ \cR (\br, \br, t) +  \cR (\br, \br, -t) } } d\br .
	\all
\end{equation}
The generalized Harada-Sasa relation~\eqref{eq:HS_modelH1} is deduced by Fourier transforming in space and time.
\vfil


\bibliography{biblio}

\end{document}